\documentclass[floatfix,twocolumn,preprintnumbers,amsmath,amssymb,aps,pra,10pt]{revtex4-2}
\usepackage{xcolor} \usepackage{xspace} \usepackage[braket, qm]{qcircuit}
\usepackage{bm, graphicx} 
\usepackage[bookmarks=false]{hyperref}

\newcommand{\Ug}{\ensuremath{U_G}}
\newcommand{\Uc}{\ensuremath{U_C}}
\newcommand{\Us}{\ensuremath{U_s}}
\newcommand{\ps}{\ensuremath{p_s}}

\usepackage{algorithm} \usepackage{algpseudocode} \usepackage{fancyhdr}
\usepackage{multirow} \usepackage{braket} \usepackage{comment}
\newcommand{\ketbra}[2]{|#1\rangle\langle#2|} 
\begin{document}

\fancyhead[R]{\ifnum\value{page}<2\relax\else\thepage\fi} \title{Analysis and
Experimental Demonstration of Amplitude Amplification for Combinatorial
Optimization} \author{Daniel Koch$^{1}$$^{*}$, Brian Pardo$^{1}$$^{2}$, Kip Nieman$^{1}$$^{2}$} \affiliation{$^{1}$Air Force Research
Lab, Information Directorate, Rome, NY } \affiliation{$^{2}$National Academy of Sciences} 
\affiliation{$^{*}$Corresponding Author: daniel.koch.13@us.af.mil  }

\begin{abstract}
Quantum Amplitude Amplification (QAA), the generalization of Grover's algorithm, is capable of yielding optimal solutions to combinatorial optimization problems with high probabilities. In this work we extend the conventional 2-dimensional representation of Grover's (orthogonal collective states) to oracles which encode cost functions such as QUBO, and show that linear cost functions are a special case whereby an exact formula exists for determining optimal oracle parameter settings. Using simulations of problem sizes up to 40 qubits we demonstrate QAA's algorithmic performance across all possible solutions, with an emphasis on the closeness in Grover-like performance for solutions near the global optimum.  We conclude with experimental demonstrations of generalized QAA on both IBMQ (superconducting) and IonQ (trapped ion) qubits, showing that the observed probabilities of each basis state match our equations as a function of varying the free parameters in the oracle and diffusion operators.
\end{abstract}

\maketitle


\thispagestyle{fancy}

\section{Introduction}


Quantum Amplitude Amplification (QAA) is the generalization of Grover's
algorithm, first proposed nearly three decades ago \cite{grover} as a
means of searching for a marked subset of an unstructured database.
This problem is framed as having access to a black-box boolean oracle
function $f:\{1,2,3,...,N\}\rightarrow \{0,1\}$  mapping elements of the
database to either $0$ or $1$, for which the solver must find the
instance(s) $f(x) = 1$ in the fewest calls to~$f$. An optimal classical
algorithm requires $O(N)$ calls to the oracle function. Encoding $f$ as
a unitary operation, Grover's algorithm solves this problem in only $O(\sqrt{N})$ calls to $f$ using a quantum computer. It
was shown shortly thereafter that Grover's algorithm is optimal
\cite{boyer,zalka}, and later that a deterministic (fixed-point) quantum
search for marked states is possible with the same speedup
\cite{long3,grover2,yoder,roy}.

Fast-forward nearly $30$ years, and the mathematical framework of
Amplitude Amplification has been analyzed and generalized in numerous
ways \cite{biham1, long2, hoyer, biham2, byrnes,kwon,szab,sun,suzuki}, used to
solve various problems beyond database searching
\cite{farhi,brassard2,childs,aaronson,janmark,anikeeva,gilliam,tezuka,durand,zhang2},
and incorporated into subroutines for other quantum algorithms
\cite{hhl,brassard1,grinko,muser}.  Many of these theoretical
advancements were developed in the absence of physical hardware, but
simultaneously there have been several successful implementations of
QAA across multiple quantum technology platforms
\cite{feng,chen,figgatt,mandviwalla,Zhang3,Zhang4,pokharel,thorvaldson,abughanem}.

In this study we focus on a particular variant of QAA which solves
combinatorial optimization problems
\cite{shyamsundar,satoh,bench,tani,zhukov,koch1,koch2}. This formulation of QAA uses an oracle operator similar to the phase-separator operator in QAOA \cite{qaoa,qaoa2,macie,blekos,bartschi,headley,bridi,xie}, applying phases proportional to every solution of a discrete cost function (Hamiltonian), which we call {\it cost oracles}. Although we
shall use the name oracle in this study to keep with normal convention,
we stress that these operators contain no black-box element to them.  The motivation for these oracle operations is their quantum circuit efficiency, addressing a common point of criticism of implementing Grover's \cite{stoud,babbush,hoefler}, namely the requirement of full $N$-Toffoli gate operators or quantum dictionaries \cite{gilliam,gilliam2} for the oracle. 

QAA as defined in this study consists of applying two alternating operations: oracle and diffusion, each containing a free parameter which can in principle vary with each application.  One way to determine these values is through measurement feedback and a classical optimizer, effectively QAOA using diffusion as the mixer \cite{bartschi,headley,bridi,xie}.  Alternatively, it has been shown that cost oracle QAA can succeed in the same manner as Grover's, using a single parameter value for each operator \cite{satoh,bench,tani,zhukov,koch1,koch2}. As in this study, using $\pi$ as the diffusion free parameter setting has been shown to be capable of achieving high probabilities of measurement for optimal solutions \cite{bench,tani,koch1,koch2} (90\%+).  

The challenge of QAA as described above is that every combinatorial optimization problem results in a unique oracle operation, which in turn requires determining the optimal free parameter setting(s) for the oracle, diffusion, and total iterations $k$.  For both regular \cite{grover} and deterministic (fixed-point) Grover's \cite{grover2,yoder,roy} optimal values for each are exactly computable, while here we show that linear cost functions (no quadratic or higher terms) produce a symmetry of solutions which can be used to derive an exact formula for predicting oracle parameter settings.  Using simulations of problem sizes up to 40 qubits we show peak probabilities and iterations $k$ for finding all possible solutions, highlighting that algorithmic performance for globally optimal solution identification becomes increasingly Grover-like with problem size.  

We conclude with experimental results showing a single iteration of generalized QAA up to 5 qubits, run on both IBMQ and IonQ, with and without noise mitigation techniques implemented by the respective vendors \cite{IBMQ,IonQ}.  Each experiment is designed around varying the free parameter in either the oracle or diffusion operator, comparing the measured probabilities of each basis state with their predicted theoretical values.  Our results show the first ever experimental demonstration of generalized QAA using cost oracles.

\section{Amplitude Amplification}\label{sec:II}

We begin by defining the QAA algorithm, given in algorithm~\ref{Alg.AmpAmp}.
\begin{algorithm}[H] \caption{Amplitude Amplification Algorithm}
\label{Alg.AmpAmp} \begin{algorithmic} \State Initialize  Qubits:
$|\Psi\rangle = |0\rangle ^{\otimes N}$ \State Prepare Equal
Superposition: $H^{\otimes N} |\Psi \rangle = |s\rangle$ \For{ $k$
iterations  } \State Apply Oracle Operation: $\Ug(\phi) |\Psi
\rangle$ or $\Uc(\ps) |\Psi \rangle$ \State Apply
Diffusion Operation: $\Us(\theta) |\Psi \rangle$ \EndFor
\State Measure $|\Psi\rangle$\end{algorithmic} \end{algorithm}

Algorithm \ref{Alg.AmpAmp} applies to both parametrized variations of QAA using Grover's oracle $\Ug(\phi)$ and cost oracles $\Uc(\ps)$.
Starting from the $N$-qubit equal superposition state $\ket{s}$ defined as
\begin{equation} 
\ket{s} = \frac1{\sqrt{2^N}}\sum_i^{2^N}\ket{Z_i}, \label{Eqn.s_state}
\end{equation}
we perform $k$ iterations of alternating oracle ($\Ug$ or $\Uc$) and diffusion ($\Us$) operations until concluding with a
measurement on $\ket{\Psi}$ in the computational basis (the Bloch sphere
$z$-axis of each qubit).  Ideally, the result of this measurement yields
the target state(s) $\ket{Z_i}$ solving the problem encoded by the
oracle. The optimal runtime of algorithm \ref{Alg.AmpAmp} using $\Ug(\pi)$ together with $\Us(\pi)$ is $k \approx \frac{\pi}{4}  \sqrt{2^N/N_m}$ for $N_m$
solutions \cite{grover,zalka,boyer}.

\subsection{Oracle operations}%

The Grover oracle $\Ug(\phi)$ is a quantum
operation defined by the unitary operator
\begin{equation} \Ug(\phi)\ket{Z_i}  = 
\begin{cases} e^{i\phi}\ket{Z_i} & \text{if } \ket{Z_i} \in \text{marked},\\ \ket{Z_i} & \text{otherwise}
\end{cases}, \label{Eqn.G_oracle}
\end{equation} which applies a phase of $e^{i \phi}$ to marked target states. The number of marked states, which we denote $N_m$, can be as few as one and typically no more than $2^N/4$ \cite{szab}.  Realizing this fully entangling operation on quantum hardware
is a known challenge \cite{stoud,babbush,hoefler}. 

We shall define $\mathbb{M}$ and $\mathbb{N}$ as sets containing all marked and non-marked $Z_i$ respectively. The Grover oracle subdivides the state $\ket{\Psi}$ into a two-dimensional subspace spanned by the orthogonal basis states in $\mathbb M$ and $\mathbb N$ as defined by
\begin{gather}
    \ket{m} \equiv \frac{1}{\sqrt{N_m}}\sum_{Z_i\in\mathbb M}\ket{Z_i}, \quad \ket{n} \equiv \frac{1}{\sqrt{N_n}}\sum_{Z_i\in\mathbb N}\ket{Z_i} \label{Eqn.M_N_sets}
\end{gather}
For an $N$-qubit
system one typically has $N_n+N_m=2^N$, but more generally this summation can
be any integer when using qudits \cite{wang} or qubits \cite{shukla}.  Writing $\ket{\Psi}$ in terms of the collective states $\ket{n}$ and
$\ket{m}$ allows the oracle operation to be rewritten as
\begin{equation}
\Ug(\phi) = \ketbra{n}{n} + e^{i\phi}\ketbra{m}{m}. \label{Eqn.G_oracle2} 
\end{equation}

The formalism of equations \ref{Eqn.M_N_sets} and \ref{Eqn.G_oracle2} can be extended to oracles which apply more than two unique phases, which is how we shall define cost oracles in this study, given by 
\begin{equation}
\Uc(p_s)\ket{Z_i} = e^{iC(Z_i)\cdot p_s}\ket{Z_i}.\label{Eqn.C_oracle}
\end{equation}
Specifically, we shall generalize the notion of the collective states $\ket{n}$ and $\ket{m}$ to $\ket{C_j}$, defined 
as
\begin{equation}
    \ket{C_j} \equiv \frac{1}{\sqrt{N_j}}\sum_{Z_i\in\mathbb C_j}\ket{Z_i},\label{Eqn.C_states}
\end{equation} 
where $C_j$ is a value obtained from evaluating a cost function C$(Z)$ using the bit string $Z_j$. Analogous to $\mathbb{N}$ and $\mathbb{M}$ in equation \ref{Eqn.M_N_sets}, each collective state $|C_j\rangle$ is defined as the summation of basis states $|Z_i\rangle$ whose bit strings are contained within the set $\mathbb{C}_j$. Specifically, each set $\mathbb{C}_j$ contains all $Z_i$ which evaluate to the same cost function value C$(Z_i) = C_j$. We can then rewrite the cost oracle as 
\begin{equation}
    \Uc(\ps) = \sum_{j}^D \ketbra{C_j}{C_j} e^{i C_j \cdot \ps}. \label{Eqn.C_oracle2}
\end{equation} The unitary operator $\Uc$ has one free parameter $\ps$, which we use to stand for the {\it phase scale} \cite{koch1,koch2} (equivalent to $\gamma$ in the QAOA literature \cite{bartschi,headley,bridi,xie}). Analogous to $\Ug$, applying $\Uc$ subdivides $\ket{\Psi}$ into a $D$-dimensional subspace spanned by the orthogonal collective states $|C_j\rangle$, where $D$ is the
number of unique evaluations obtainable from C$(Z)$. Importantly, the quantum circuit construction of $\Uc$ does not require any knowledge of these collective states or the $|Z_i\rangle$ contained within them (see \cite{koch1,koch2} and appendix \ref{sec:appendixE}), which are assumed to be unknown to the solver.

To conclude, $D=2$ is by
definition a Grover oracle, while $D \in [3,2^N]$ defines cost oracles, where $D$ is problem dependent on C$(Z)$. Note that equations \ref{Eqn.C_states} and \ref{Eqn.C_oracle2} reduce to the form of \ref{Eqn.M_N_sets} and \ref{Eqn.G_oracle2} for the case $D=2$,
$\ps=\phi$, $\{ |C_1\rangle, |C_2\rangle \} = \{ |n\rangle,
|m\rangle \}$, $\{ C_1, C_2 \} = \{ 0, 1 \}$, $\{ N_1, N_2 \} = \{ N_n, N_m
\}$.  Thus, our equations for cost oracles are a generalization of Grover's to higher dimensions $D>2$.

\subsection{Diffusion}%

\begin{figure}[H] \centering 
\input{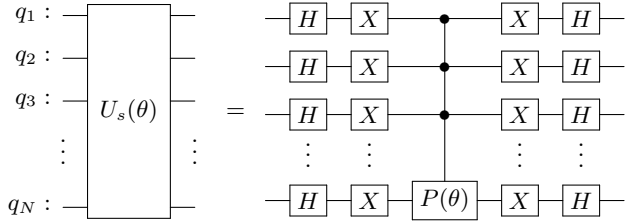}
\caption{ Quantum circuit for the diffusion
operator $\Us(\theta)$ acting on $N$ qubits.} 
\label{Fig.Dif_qc}
\end{figure}

We now turn to the diffusion operator $\Us(\theta)$, defined by
\begin{equation}
\Us(\theta) =
\mathbb{I} - ( 1 - e^{i \theta} ) \ketbra{s}{s},  \label{Eqn.Diff} 
\end{equation} corresponding to the quantum circuit construction shown in figure \ref{Fig.Dif_qc}.  This is the standard construction of $\Us(\theta)$ as originally proposed by Grover \cite{grover}, also referred to as the Grover mixer in QAOA 
\cite{blekos,headley,bartschi}. 
This multiqubit operation is the $N$-qubit equivalent to $P(\theta)$ shown in equation \ref{Eqn.phase_gate}, applying a phase of $e^{i \theta}$ to the basis state $|1\rangle^{\otimes N}$, and no phase to all other $|Z_i\rangle$.
\begin{equation}
    P(\theta) = \begin{bmatrix}
        1 & 0 \\ 0 & e^{i\theta}
    \end{bmatrix} \label{Eqn.phase_gate}
\end{equation}

Because oracle operations only apply phases, in QAA the diffusion operator is solely responsible for increasing and decreasing amplitude magnitudes (probabilities).  The manner in which it does so is proportional to $\bar{\alpha}$, the arithmetic mean of all amplitudes $\alpha_i$ (amplitudes of each basis state: $\alpha_i = $ $\langle Z_i|\Psi\rangle$), given by 
\begin{align}
    \bar{\alpha}  &= \frac{1}{\sqrt{2^N}} \braket{s|\Psi} = \frac{1}{2^N} \sum_{i=1}^{2^N} \braket{Z_i|\Psi}
 \label{Eqn.abar1} \\
 &= \frac{1}{2^N} \sum_{i=1}^{D} N_i \cdot \alpha_i, \label{Eqn.abar2}
 \end{align}
 while
 \begin{equation}
     \langle Z_j| \Us(\theta)|\Psi \rangle = \alpha_j - (1-e^{i \theta})\bar{\alpha}. \label{Eqn.Diff_amp}
 \end{equation}
 
 Equation \ref{Eqn.Diff_amp} is the change in amplitude experienced by each basis state $|Z_i\rangle$ resulting from diffusion.  The typical choice for diffusion is $\theta = \pi$, which from equation \ref{Eqn.Diff_amp} results in $\Delta \alpha_i = \alpha_i - 2 \bar{\alpha}$.  This change in amplitude is then maximal when $\alpha_i$ and $\bar{\alpha}$ are $\pi$ phase different, which is the situation created by $\Ug(\pi)$ for the marked state(s) \cite{grover}. Next in section~\ref{sec:III} we show that this same $\pi$ phase matching condition can be achieved for $\Uc$ encoding linear cost functions.

\section{ Solving Linear Cost Functions } \label{sec:III}  %

For $\Ug$, the optimal $\phi$, $\theta$, and $k$ values for Algorithm \ref{Alg.AmpAmp} are determined by the size of the unmarked and marked subspaces $N_n$ and $N_m$ \cite{grover,yoder,roy}. For cost oracles $\Uc$, these parameters are determined by $C_i$ and $N_i$.  In practice however, these values are unknown except through evaluations C$(Z)$, so a realistic implementation of QAA for combinatorial optimization needs to be able to determine $\ps$, $\theta$, and $k$ values without excessive classical precalculation. Here we demonstrate one such example for linear cost functions and discuss the mathematical properties which make it possible.

\begin{figure*}
\centering 
\includegraphics[scale=.22]{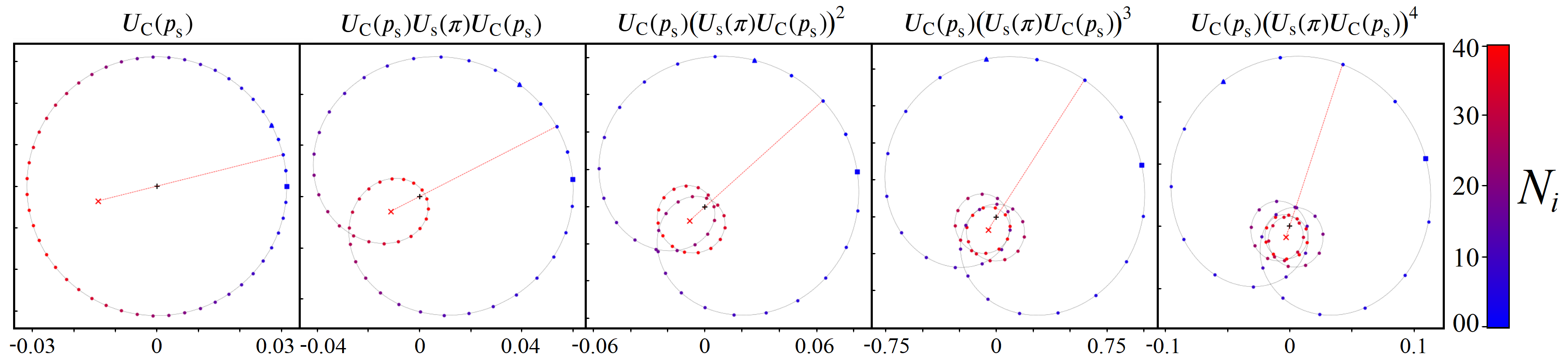} \caption{ Five complex plane plots of $|\Psi\rangle$ after the
application of the operators shown above each panel, corresponding to the
10-qubit C$(Z)$ given in appendix \ref{sec:prob_instance}.  The value of $\ps$ used in
$\Uc$ is from equation \ref{Eqn.ssp_ps_opt3}, for $C_i=2$.  Each
colored point in the plots represents the amplitudes contained within one collective
state $|C_i\rangle$, with $N_i$ indicated by the accompanying color scale. The extrema states $|C_{\textrm{min}}\rangle$ and $|C_{\textrm{max}}\rangle$ are plotted as
square and triangle markers respectively.  In each panel the value of
$\bar{\alpha}$ is depicted by the red $\times$, and the origin of the complex plane
with a black +.  A red-dotted line is drawn in each panel from $\bar{\alpha}$ to the
collective state $|C_i=2\rangle$,
illustrating their $\pi$ phase difference via the line's intersection with the origin.}
\label{Fig.Q10_ssp_cp} 
\end{figure*}

\subsection{Problem
Symmetry and \ensuremath{\bar{\alpha}}}

Consider the case of a set of $N$ randomly selected, real-valued weights $W_i$ with $\mathbb{W} = \{ W_1, W_2, ... , W_N  \}$.  We define a linear cost function as
\begin{equation} 
\textrm{C}(Z_i) = \sum_{i=1}^N W_i \cdot
z_i, \label{Eqn.ssp_cost_function2} 
\end{equation}
where each binary variable $z_i$ is assigned to one weight $W_i$.  Implementing C$(Z)$ according to equation \ref{Eqn.ssp_cost_function2} as a cost oracle $\Uc(\ps)$ is achieved through single qubit phase gates $P(W_i\cdot \ps)$ on each qubit \cite{koch1,koch2} (appendix \ref{sec:appendixE}). 

The key feature of this C$(Z)$ that allows one to predict optimal $\ps$ values is the symmetric distribution of the solution space about the arithmetic mean $\bar{C}$, given by
\begin{eqnarray} 
\bar{C} &=& \frac{1}{2^N} \sum_i^{2^N} C_i
\label{Eqn.c_bar1} \\ &=&  \frac{\textrm{C}(Z_j) + \textrm{C}(\neg Z_j)}{2}
\label{Eqn.c_bar2} .
\end{eqnarray}  
Equation \ref{Eqn.c_bar1} is generic to all C$(Z)$, while
\ref{Eqn.c_bar2} is a property of linear C$(Z)$ as defined by equation \ref{Eqn.ssp_cost_function2} (proof in appendix \ref{sec:appendixB}).  For linear C$(Z)$ the value of $\bar{C}$ can be obtained by evaluating any $Z_j$ and its bitwise inverse $\neg Z_j$. The significance of this symmetry of $C_i$ solutions about $\bar{C}$ is that the resulting mean amplitude after the first oracle operation $\Uc(p_{\textrm{s}})|s\rangle$ is
\begin{equation} 
\bar{\alpha} = |\bar{\alpha}| e^{i \bar{C} \cdot
\ps},  \label{Eqn.abar_cbar} 
\end{equation} 
whereby the phase of $\bar{\alpha}$ is proportional to $\bar{C}$ (appendix \ref{sec:appendixB}). Even though the value of $\bar{\alpha}$ itself requires complete information of all $|C_i\rangle$, for linear C($Z$) knowing the phase of $\bar{\alpha}$ does not. Thus, using equation \ref{Eqn.abar_cbar} we can derive a formula for $\ps$ such that the
phase of $\bar{\alpha}$ is made to be exactly $\pi$ different from the phase applied to
any target collective state $|C_i \rangle$ of one's choosing.

\begin{eqnarray} \bar{C} \cdot \ps - C_i \cdot \ps   =
\pm \pi, \label{Eqn.ssp_ps_opt1}
\\ \ps = \frac{ \pm  \pi}{
\bar{C} - C_i }  \hspace{0.25cm} \textrm{for}
\hspace{0.1cm} C_i \neq \bar{C} \label{Eqn.ssp_ps_opt3} \end{eqnarray} 

Equation \ref{Eqn.ssp_ps_opt3} ensures a $\pi$ phase difference between $\bar{\alpha}$ and the target state $|C_i\rangle$ after the first oracle operation, while in appendix \ref{sec:appendixB} we show that this phase difference holds for subsequent iterations when using $\Us(\pi)$.  A visualization of this property is shown in figure \ref{Fig.Q10_ssp_cp}, displaying a series of complex plane plots for $|\Psi\rangle$ after applying $\Uc(\ps)$ according to equation
\ref{Eqn.ssp_ps_opt3} for the first five iterations, corresponding to the
10-qubit C$(Z)$ in appendix \ref{sec:prob_instance}.  As indicated by the axes scale below each plot, the probability of measuring the target state is increasing with each subsequent iteration in a manner exactly analogous to Grover iterations $\Us(\pi) \Ug(\pi)$ ($\pi$ phase between $|m\rangle$ and $\bar{\alpha}$), but for an oracle encoding a cost function $\Us(\pi) \Uc(\ps)$.

\subsection{Algorithmic Performance} 

The success of QAA for solving a combinatorial optimization problem encoded as $\Uc$ is dependent on correctly choosing the three parameters $\ps$, $\theta$, and $k$. A reliable means of determining these parameters for problems such as QUBO or harder \cite{satoh,bench,tani,zhukov,koch1,koch2} remains an open research question, while here shall show that equation \ref{Eqn.ssp_ps_opt3} for $\ps$ together with $\Us(\pi)$ is sufficient for reliable algorithmic performance. 

\begin{figure} \centering \includegraphics[scale=.22]{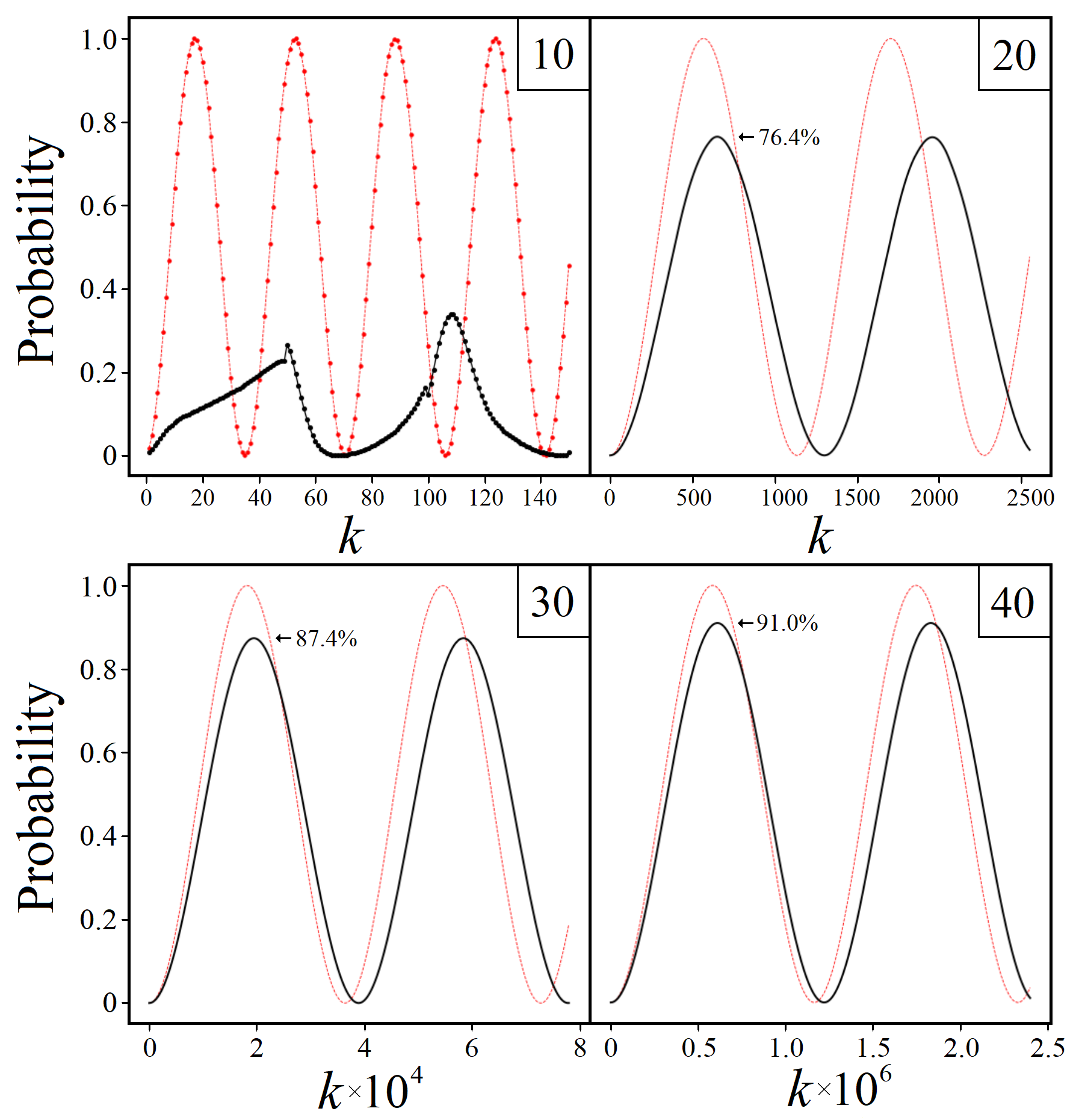} 
\caption{Simulations of QAA showing probability versus iterations $k$ for qubit sizes $N=10,20,30,40$, comparing the joint probability of $|C_i=2\rangle$ and its inverse (solid-black lines) to $|m\rangle$ for standard Grover's (dashed-red lines) for $N_m=2$. The $p_{\textrm{s}}$ value used for each $\Uc$ is from equation \ref{Eqn.ssp_ps_opt3}, while the weights of each linear C($Z$) are $\mathbb{W}_1$ from appendix \ref{sec:exp_states}. } \label{Fig.ssp_prob_plots}
\end{figure}

For the 20-qubit linear C($Z$) composed of the integer weights $\mathbb{W}_2$ given in appendix \ref{sec:exp_states}, figure \ref{Fig.solving_for_T} shows the joint peak probabilities (solid-colored lines) obtainable for the ten most optimal $|C_i\rangle$ states as a function of $\ps$. Specifically, each colored curve represents the combined probability of measuring  $|C_i=T\rangle$ or its inverse $|C_i=2\bar C - T\rangle$ (for example, $|C_i=-222\rangle$ and $|C_i=194\rangle$ shown in blue). Over the range of $p_{\textrm{s}}$ values shown, each peak probability was obtained by simulating algorithm \ref{Alg.AmpAmp} up to the iteration $k$ where it peaks for the first time (see figure \ref{Fig.ssp_prob_plots}). The resonance-like shape of these curves is a well-understood feature of Grover's algorithm \cite{long, long2, hsieh, romanelli_quantum_2006}, also observed for harder combinatorial problems such as QUBO \cite{koch1,koch2},  and emerges from the condition for constructive interference for the target state (see appendix \ref{sec:appendixD} for more details).

\begin{figure}\centering \includegraphics[scale=.31]{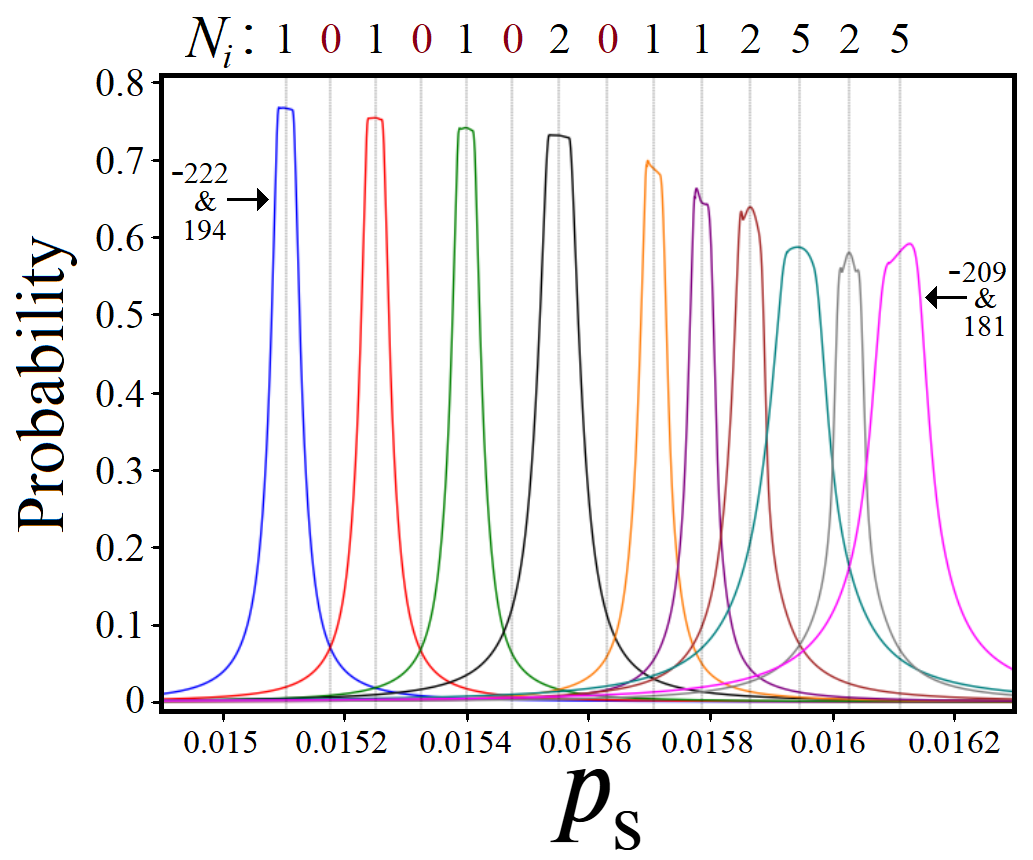} \caption{(solid-color lines) Combined peak probability as a function of
$\ps$ for the ten highest (lowest) $|C_i\rangle$ states, for the $N=20$ linear C$(Z)$
composed of the weights $\mathbb{W}_2$ from appendix \ref{sec:exp_states}. (gray-dashed lines) The $\ps$ values produced from equation \ref{Eqn.ssp_ps_opt3} for the integers $ T \in [-222,-209]$, with $N_i$ values reported atop each line. }\label{Fig.solving_for_T}
\end{figure}

Accompanying the curves in the figure are vertical gray lines corresponding to $\ps$ values obtained from equation \ref{Eqn.ssp_ps_opt3} for the integers $ T \in [-222,-209]$. Above each line is the corresponding $N_i$ value for each $|C_i = T\rangle$ state (or zero if no such solution exists). As evidenced by the closeness of the lines to each of the ten probability curve peaks, linear C($Z$) with integer weights is the ideal scenario for QAA using equation \ref{Eqn.ssp_ps_opt3}. Specifically, a solver knows to only check integer values for equation \ref{Eqn.ssp_ps_opt3}.   Although it is a simple problem instance, it demonstrates that at 20 qubits an exact equation for $\ps$ is capable of producing measurement probabilities of $60$-$75$\%+ for $|C_i\rangle$ near the extrema. However, these probabilities are low compared to larger qubit problem sizes, such as those shown in figure \ref{Fig.ssp_prob_plots}.

\begin{figure} \centering \includegraphics[scale=.34]{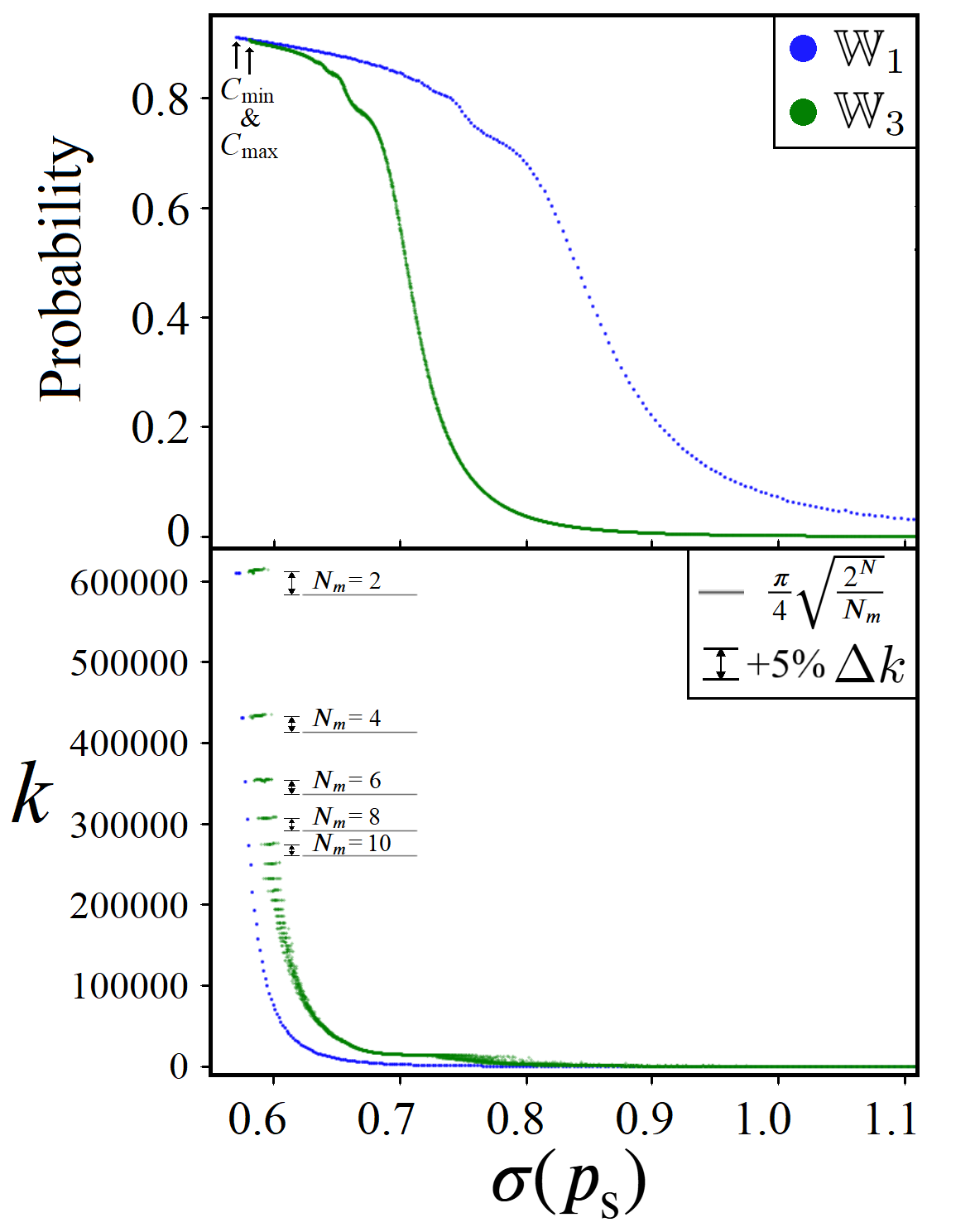} \caption{(top) Plot of peak achievable probabilities for each $|C_i\rangle$ state (colored circles) as a function of
$\sigma(\ps)$ using $\ps$ from equation \ref{Eqn.ssp_ps_opt3}, for the $N=40$ cases $\mathbb{W}_1$ and $\mathbb{W}_3$ given in appendix \ref{sec:exp_states}. (bottom) Plot of iterations $k$ corresponding to the probabilities shown in the top plot. For comparison, lines indicating the number of iterations for Grover's ($\Ug(\pi)$) using various $N_m$ are plotted. Beside each line is $\Delta k$, showing a 5\% increase in $k$ from Grover's.  } \label{Fig.Q40_prob_plots} \end{figure}

Figure \ref{Fig.ssp_prob_plots}, together with figure \ref{Fig.Q40_prob_plots} illustrate two key features of QAA algorithmic performance. Firstly, as qubit size increases the achievable probabilities of solutions near the extrema, and the iterations $k$ to reach these probabilities both become increasingly Grover-like. Secondly, within a single problem instance the peak achievable probabilities of each $|C_i\rangle$ decrease for solutions further away from the extrema. This first feature is an important strength for cost oracle QAA, illustrating $\Uc(\ps)$ is capable of producing comparable probabilities to $\Ug(\phi)$ without the need for expensive oracle circuits \cite{stoud,babbush,hoefler,gilliam,gilliam2}, especially for extrema $C_i$ which are typically the solutions of interest for combinatorial optimization problems. 

Focusing now on figure \ref{Fig.Q40_prob_plots}, the top plot shows the complete spectrum of achievable probabilities for all $|C_i\rangle$ for two $40$-qubit linear C($Z$) using $\ps$ from equation \ref{Eqn.ssp_ps_opt3}, while the bottom plot shows the corresponding number of iterations $k$ needed to achieve these probabilities. For completeness, given below in equation \ref{Eqn.sigma_cz} is $\sigma(\ps)$, the standard deviation of all $C_i$ solutions after scaling by $p_{\textrm{s}}$, which is a convenient way to plot all $C_i$ solutions for both C($Z$) on the same axis.

\begin{equation} 
\sigma(\ps) =  \sqrt{ \frac{\sum_i^{2^N} (C_i\cdot \ps -
\bar{C}\cdot \ps)^2}{2^N} }  \label{Eqn.sigma_cz}
\end{equation} 

The significance of figure \ref{Fig.Q40_prob_plots} is the bottom plot, particularly the black lines which show the value of $k$ for standard Grover's for various values of $N_m$.  Beside each black line is a 5\% increase in $k$, labeled $\Delta k$, illustrating that for these $40$-qubit problem instances the optimal number of iterations for cost oracle QAA is consistently around $5$\% more than standard Grover's. Thus, the results shown in figures \ref{Fig.solving_for_T} - \ref{Fig.Q40_prob_plots} illustrate a special case of QAA, oracles encoding linear C($Z$) composed of integer weights, whereby a solver can reliably determine near optimal parameter settings for $\ps$, $\theta$, and $k$.  These results are particularly encouraging because the same underlying features have been observed for QAA using oracles encoding QUBO \cite{satoh,bench,tani,zhukov,koch1,koch2}, suggesting that these harder problem instances may also have reliable strategies for approximating optimal parameter values as well.

\section{ Experimental
Demonstration }\label{sec:IV}   %

To conclude this study here we present experimental results demonstrating current hardware progress towards the realization of generalized QAA \cite{kwon} on both IBMQ's superconducting qubits as well as IonQ's trapped ion qubits. Each experiment is a single iteration of $\Us(\theta)\Uc(\ps)$ or $\Us(\theta)\Ug(\phi)$, varying either $\theta$ or $\ps$ across $100$ different values and tracking the measured probabilities of each basis state $|Z_i\rangle$. At each parameter value we ran the quantum circuit $10,000$ times, totaling one million circuit executions per experiment per qubit size.  Our results confirm that these probabilities match the theoretically predicted values from our equations, to our knowledge the first ever such experimental demonstration of generalized QAA using cost oracles.

In total we conducted three unique experiments, which we refer to by number, with quantum circuits for each experiment given in appendix \ref{sec:appendixE}. Each experiment was implemented for qubit sizes $N=2,3,4,5$, run on five unique devices: IonQ's aria1 ion trap, IBMQ's fez and torino (heron), and IBMQ's kyiv, and brisbane (eagle) superconducting processors, with and without error mitigation techniques provided by the respective hardware vendors \cite{IBMQ,IonQ}. The implementation of the multicontrolled phase gate $C^N$-$P(\theta)$ used for both $\Us(\theta)$ and $\Ug(\pi)$ is adapted from \cite{barenco,schuchthesis}, discussed in detail in appendix \ref{sec:appendixE}.

\subsection{The First Iteration}%

Here we present equations for the theoretical quantum states of experiments 1-3, corresponding to the first
iteration of QAA using Grover and cost oracles, analogous to
\cite{kwon,roy}.  We begin with equations \ref{Eqn.avg_binary} and \ref{Eqn.avg_non_binary} below 

\begin{eqnarray} \bar{\alpha}(\phi) &=&  \Big(  \frac{1}{2^N}
\Big)^{\frac{3}{2}}  \cdot  \Big( N_n +  e^{i \phi} N_m \Big)
\label{Eqn.avg_binary}, \\ \bar{\alpha}(\ps) &=&  \Big(
\frac{1}{2^N} \Big)^{\frac{3}{2}} \cdot \sum_j^D  e^{i C_j \cdot
\ps} N_j  \label{Eqn.avg_non_binary} ,
\end{eqnarray} 

which are the average amplitudes $\bar{\alpha}$ (equation \ref{Eqn.abar2}) as a function of $\phi$ and $\ps$ for the states $\Ug(\phi)|s\rangle$ and $\Uc(\ps)|s\rangle$ respectively.
Using these equations in combination with \ref{Eqn.Diff_amp}, we obtain the quantum
states after the first iteration for both oracle types given by

\begin{align} \Us(\theta) \Ug (\phi)  | s \rangle
= \sqrt{N_n} \biggl( \frac{1}{\sqrt{2^N}} - \bar{\alpha}(\phi)( 1 -
e^{i\theta} )  \biggr)|n\rangle \nonumber \\ + \sqrt{N_m} \biggl( \frac{e^{i
\phi}}{\sqrt{2^N}} - \bar{\alpha}(\phi)( 1 - e^{i\theta} )  \biggr) |m\rangle,
\label{Eqn.first_k_binary} \\
\Us(\theta) \Uc (\ps)  |
s \rangle = \hspace{2cm} \hspace{2.7cm} \nonumber \\ \sum_j^D \sqrt{N_j}
\biggl( \frac{e^{i C_j \cdot \ps}}{\sqrt{2^N}} -
\bar{\alpha}(\ps)( 1 - e^{i\theta} )  \biggr) |C_j\rangle.
\hspace{0.3cm} \label{Eqn.first_k_non} 
\end{align}

Experiments 1 and 2 correspond to equation \ref{Eqn.first_k_non}, whereby in exp. 1 we prepared and measured 100 quantum circuits for the state $\Us(\pi)\Uc(p_{\textrm{s}})$ (varying the parameter $\ps$) and likewise $\Us(\theta)\Uc(1)$ for exp. 2 (varying $\theta$). Experiment 3 corresponds to states of equation \ref{Eqn.first_k_binary} $\Us(\theta)\Ug(\pi)$, $N_m=1$, varying $\theta$ over the full range of $2\pi$.  Taking the squared magnitude of the amplitudes given in equations
\ref{Eqn.first_k_binary} and \ref{Eqn.first_k_non} yields the expected
theoretical measurement probabilities for each collective state, which we experimentally verify in the coming subsections.

\subsection{Performance Metric}

\begin{figure}[h]\centering \includegraphics[scale=.3]{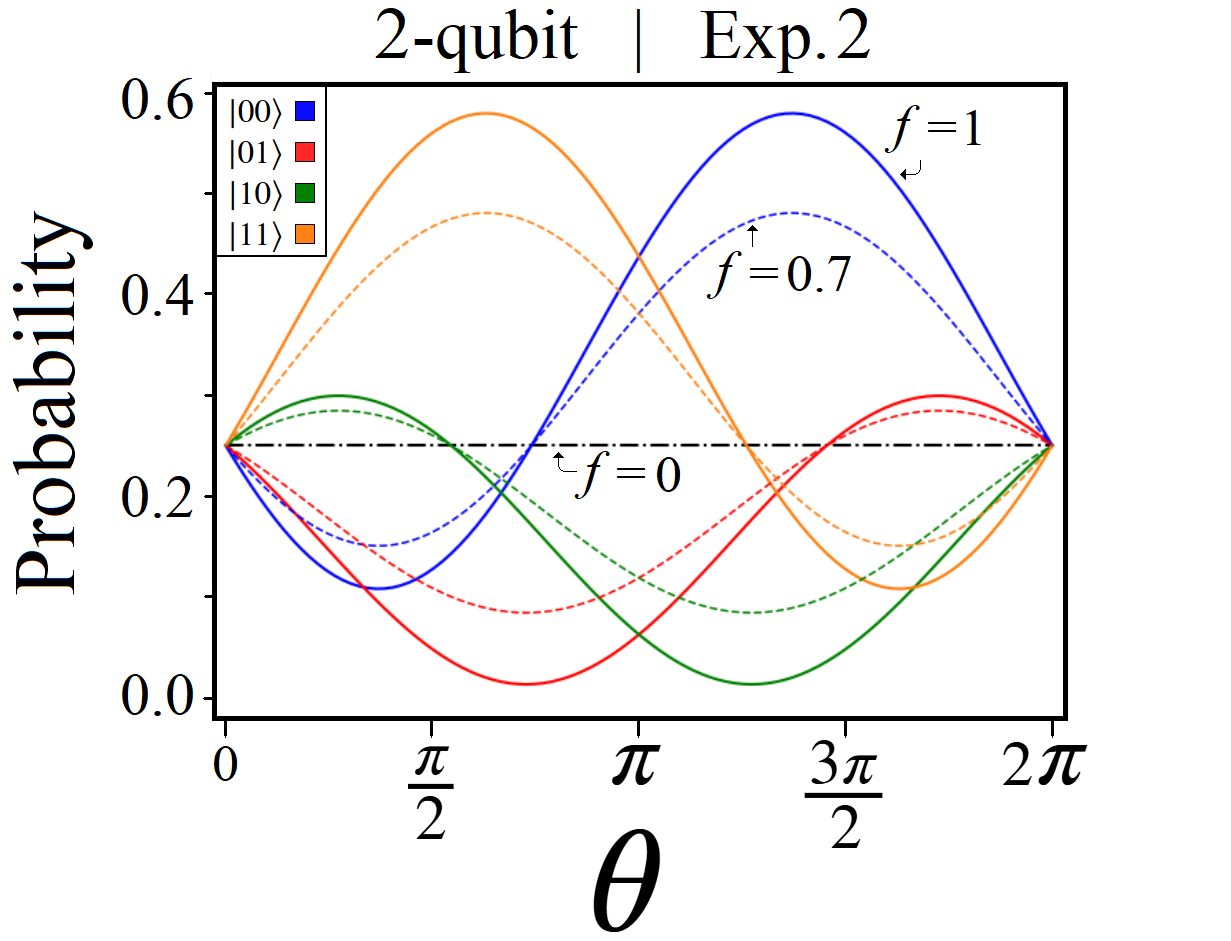} \caption{ (solid-colored) Probabilities representing $f=1$ for
each basis state as a function of
diffusion angle $\theta$ for experiment 2, $N=2$, as predicted from
equation \ref{Eqn.first_k_non} over the range $\theta \in [0,2 \pi]$. For comparison,
probabilities corresponding to $f=0.7$ (dashed-colored) and
$0$ (black-dash-dotted) are also plotted.  } \label{Fig.f_metric_plot}
\end{figure}

Because each experiment is a composition of 100 different quantum
circuits, here we shall briefly describe the quantity $f$, which is a metric that pools the results from all 100 circuits into a single value comparing experimentally measured probabilities versus theory, full mathematical description given in appendix \ref{sec:f_metric}. The quantity $f$ is a ratio comparing experimentally observed measurement counts of each $|Z_i\rangle$ to probabilities predicted by equations \ref{Eqn.first_k_binary} and \ref{Eqn.first_k_non}: $f=1$ means that the basis state was observed with the same probability as theoretically expected, while $f=0$ means an observed probability of $1/2^N$ (the equal superposition state $|s\rangle$), i.e. a completely decohered quantum state. Values of $f<0$ occur when a basis state's measured probability was found to be on the opposite side of $1/2^N$ as predicted from theory, such as in figure \ref{Fig.Notables2} to come, signaling that the qubits had reached an unintended final $|\Psi\rangle$ not attributable to decoherence. Shown in figure \ref{Fig.f_metric_plot} is an example plot illustrating three theoretical values of $f$ (no experimental data) for experiment 2, the quantum state $|\Psi\rangle = \Us(\theta)\Uc(1)$ over the full $2\pi$ range of $\theta$.

\subsection{Improvement from Error Mitigation}%

\begin{figure}[h]\centering \includegraphics[scale=.3]{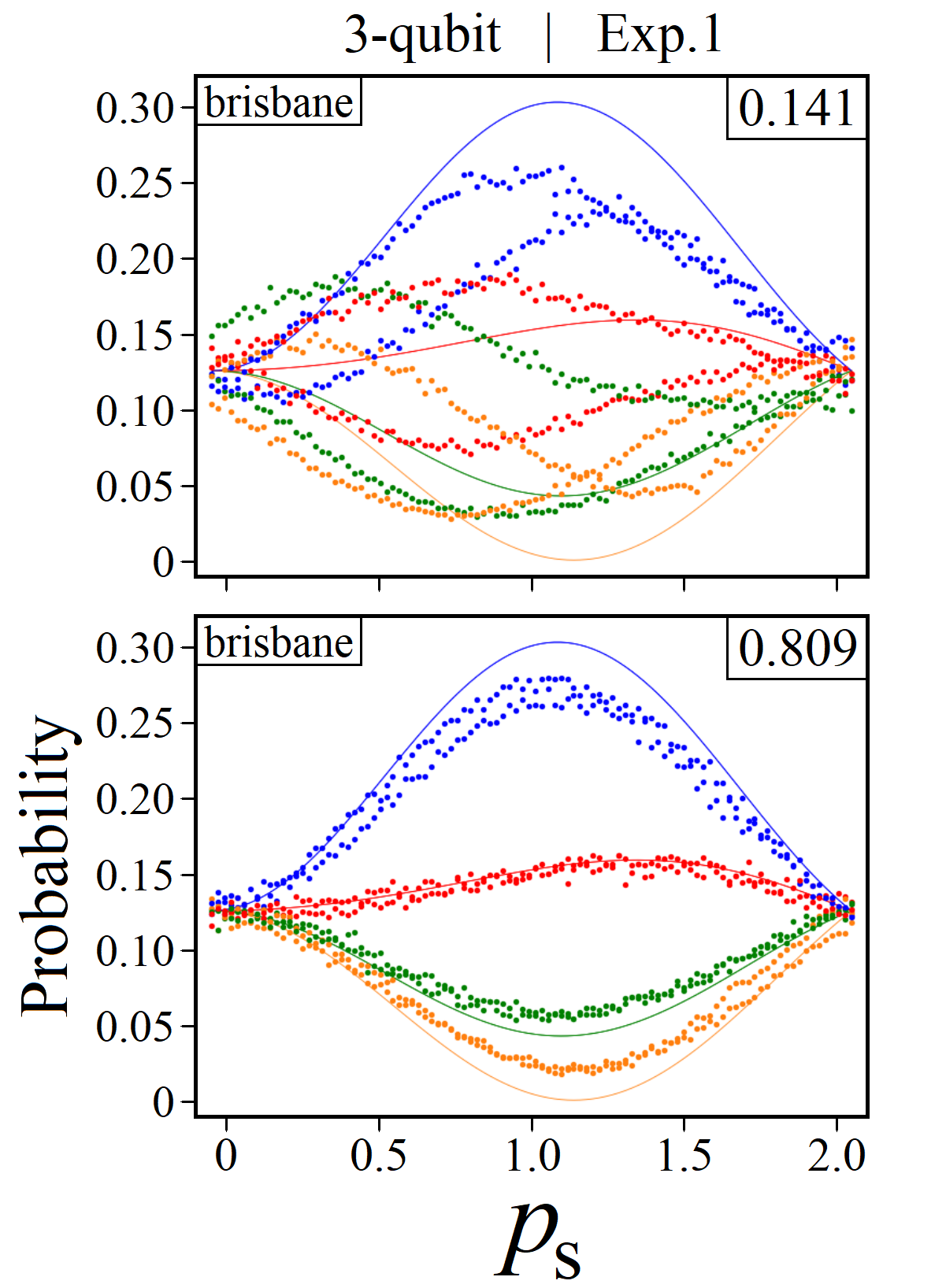}
\caption{ A comparison of measurement results for exp. 1, $N=3$, on IBMQ's brisbane with (bottom) and without (top) error mitigation. The solid-colored lines represent probabilities for $f=1$ (equation \ref{Eqn.first_k_non}), while the colored data
points are the experimental measurement probabilities obtained for each basis state. Reported in the top right corner is the cumulative $f$ for the experiment, $f_{\textrm{exp}}$ from equation \ref{Eqn.f_exp} in appendix \ref{sec:f_metric}. } \label{Fig.Notables2} \end{figure}

A recent advancement in both IBMQ and IonQ's commercially available quantum computing platforms is the addition of error mitigation as an option for users \cite{IBMQ,IonQ}.  More specifically each service offers both Pauli gate twirling and dynamical decoupling.  In this study we report on qubit performance for error mitigation `on' and `off', which refers to the use of both techniques together by the respective hardware vendors, or neither.  An example of difference in performance for on versus off is illustrated in figure \ref{Fig.Notables2}. In each plot the solid-colored lines represent $f=1$ (equation \ref{Eqn.first_k_non}), while the colored data points show the measured probability counts for each $|Z_i\rangle$ state (colored according to the $|C_i\rangle$ each basis state belongs to), with $f_{\textrm{exp}}$ for the entire experiment reported in the top right ($f_{\textrm{exp}}$ given in appendix \ref{sec:f_metric}). Examples of $f < 0$ for individual $|Z_i\rangle$ can be seen in the top plot (error mitigation off), illustrating noisy $|\Psi\rangle$ states with clear underlying structure (ripe for error mitigation / correction techniques) as opposed to complete decoherence.

\subsection{Experimental Results}%

\begin{figure}[h]\centering \includegraphics[scale=.24]{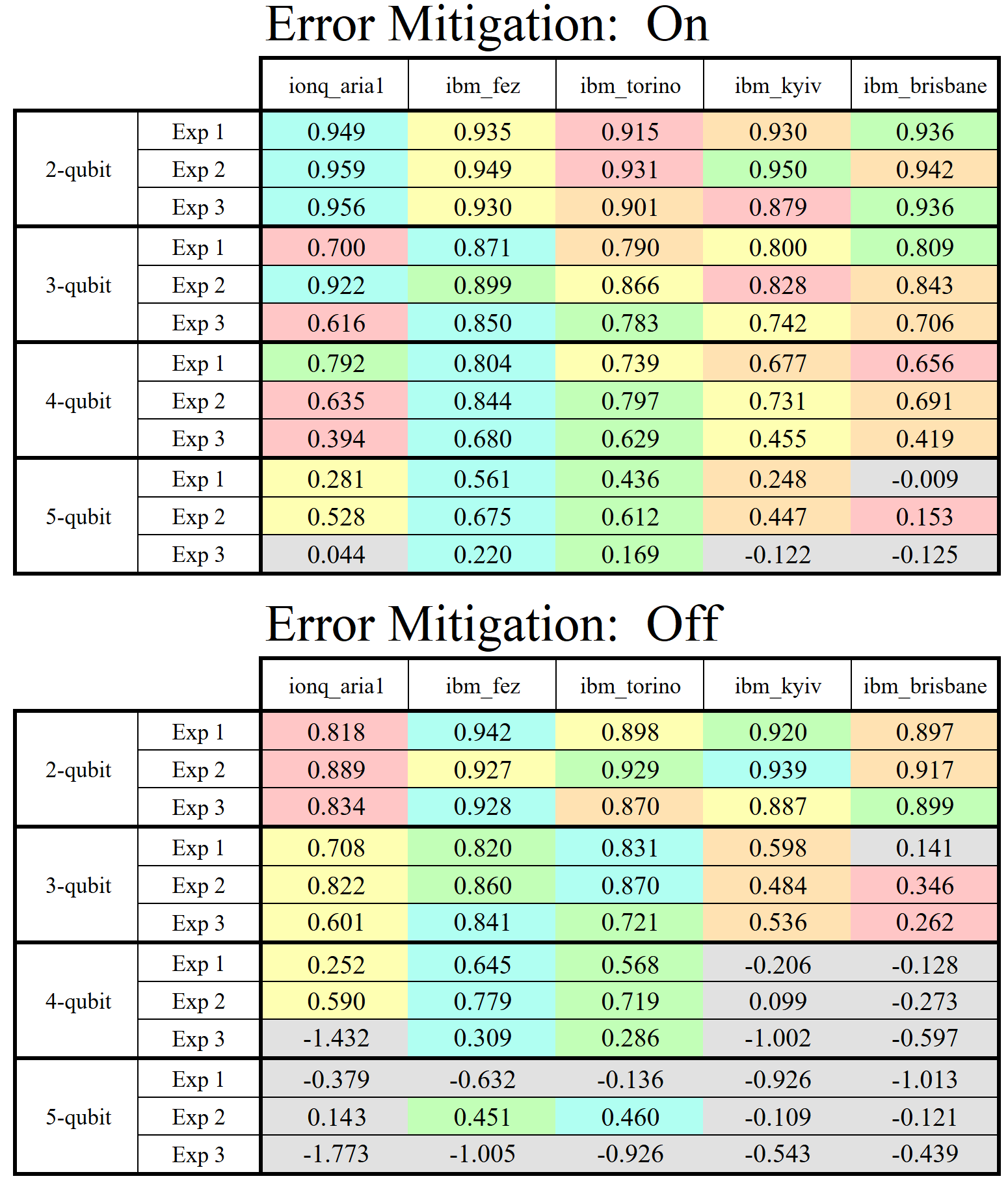} \caption{ Two tables summarizing the experimental results
of exps. 1 - 3 for error mitigation on (top) and off (bottom).  The numerical value of each cell is $f_{\textrm{exp}}$ from
equation \ref{Eqn.f_exp}, with colors indicating ranked performance. Gray
cells indicate values less than $0.15$.    } \label{Fig.Exp1_2_3_results}
\end{figure}

Given in figure \ref{Fig.Exp1_2_3_results} below are the full results for exps. 1 - 3, with error mitigation techniques
turned on (top) and off (bottom).  The numerical values shown in each table
are $f_{\textrm{exp}}$ from equation \ref{Eqn.f_exp} (appendix \ref{sec:f_metric}), the average $f$ value from all $2^N$ basis states.  The color of each
cell indicates the ranked performance of each device within a single
experiment, from best to worst: [blue, green, yellow, orange, red].  Cells
that are instead colored gray are for values $f_{\textrm{exp}} < 0.15$. 

\begin{figure*}\centering \includegraphics[scale=.26]{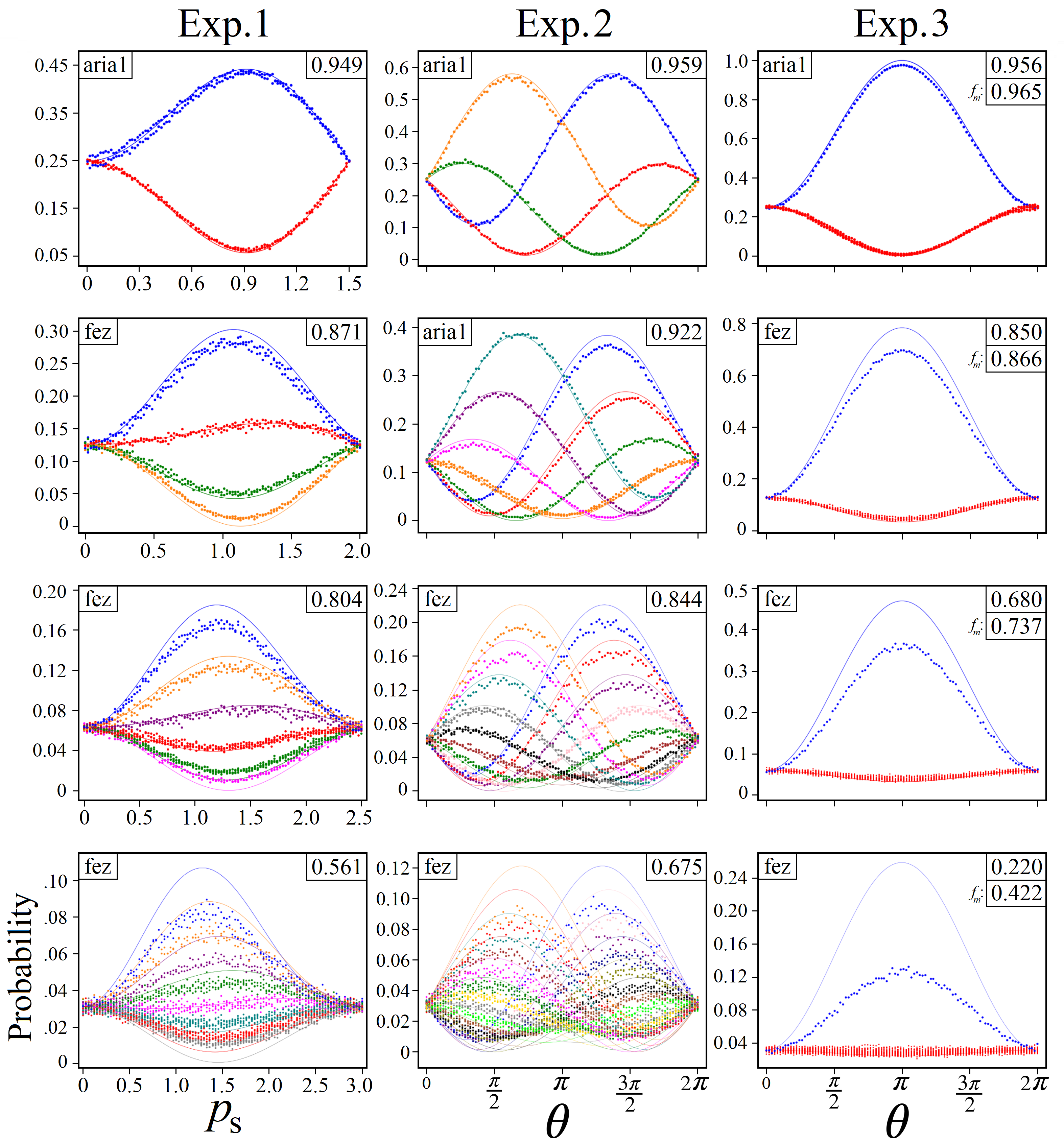} \caption{ Experiment results showing the highest $f_{\textrm{exp}}$ achieved for each combination of experiment and qubit size. The data points in each plot are Meas.($|Z_i\rangle$) from equation \ref{Eqn.Meas}, colored according to $|C_i\rangle$ states (see appendix \ref{sec:exp_states}).  The solid-colored lines in each plot represent $f=1$ for each collective state, obtained from equations \ref{Eqn.first_k_binary} and \ref{Eqn.first_k_non}. Reported in the top right of each plot is $f_{\textrm{exp}}$, as well as $f_{m}$ for exp. 3 ($f$ for the blue data points corresponding to $|m\rangle$, the all $|1\rangle$ state). }
\label{Fig.Top_Results} \end{figure*}

A summary of the results shown in figure \ref{Fig.Exp1_2_3_results} is as
follows.  The effect of error mitigation provided by both vendors improved
$f_{\textrm{exp}}$ in 55 out of the 60 experiment runs, producing an overall
average increase in $f_{\textrm{exp}}$ by $0.39$. Regarding the performance
of the individual devices, IBMQ's heron processors fez and torino ranked
first and second in 13 out of 18 experiments for qubit sizes $N \geq 3$.
However, with error mitigation IonQ's aria1 trap was first in all three
2-qubit experiments, including the highest overall $f_{\textrm{exp}}$ value
of $0.959$ as well as the highest 3-qubit value of $0.922$.  Shown in figure \ref{Fig.Top_Results} are the highest $f_{\textrm{exp}}$ achieved for all experiments and qubit sizes.

\clearpage

\section{Conclusion}

In this study we have generalized the the common $|n\rangle$ and $|m\rangle$ collective state formalism of Grover's algorithm to cost oracles encoding combinatorial optimization problems \cite{satoh,bench,koch1,koch2,tani,zhukov}, and shown that for the special case of linear C($Z$) an exact equation exists for determining the free parameter value of the oracle operator.  In section III we used simulations of QAA up to 40 qubits to demonstrate the closeness in algorithmic performance of cost oracles versus Grover oracles, as well as the range of algorithmic capability QAA using cost oracles can provide over all possible solutions. And finally, in section IV we verify our derived equations of generalized QAA \cite{kwon,roy} on two different qubit technologies, showcasing progress in state-of-the-art commercial quantum devices and their respective error mitigation capabilities.

\subsection{Outlook and Future Research}

The motivation for studying cost oracle QAA \cite{satoh,bench,tani,zhukov,koch1,koch2} stems from the quantum circuit efficiency of $\Uc$ \cite{mckay,pelofske} as compared to $\Ug$ \cite{gilliam,gilliam2,zhang2,Zhang3,Zhang4,stoud,babbush,hoefler}.  Progress in reducing circuit depth \cite{yong,dasilva,dreier,nie,silva2} combined with error correction \cite{google1,google2} is one solution towards achieving large $N$-Toffoli gates, but it is also possible that the technology may go beyond 2-qubit gates \cite{maslov,kim,warren,katz} to reduce circuit depth.  There is
also the potential for realizing QAA outside of the gate-based model, such
as quantum programmable processors \cite{harris,mohit} for implementing the oracle
and diffusion operators as photonic integrated circuits.

Much of this work focuses on oracles encoding linear cost functions composed of integer weights to demonstrate ideal conditions for cost oracle QAA. Understanding these conditions for success are important because similar studies have shown that the same algorithmic performance can be achieved for QUBO and more complex problems \cite{bench,zhukov,koch2,koch1,tani}, but determining $\ps$ (the free parameter setting of the oracle) remains an open research question. Alternatively, there is the Grover-Mixer QAOA  \cite{bartschi,headley,bridi,xie} style of determining optimal parameter settings for each iteration via measurement results and a classical optimizer. The advantage of QAA as demonstrated in section III is efficiency, requiring ideally only a few measurements to find the desired solution of a cost function. If QAA can be supplemented with precomputed optimal operator settings, ideally exact but also possibly approximate, then quantum's computational merit becomes how quickly (physical time) QAA goes from $|s\rangle$ to measuring $|Z_i\rangle$.

Lastly, the experimental results of
section IV show current commercial hardware's ability to achieve QAA up to $N=5$ qubits, which is really to say their ability to implement the $N$-Toffoli gate operation. Based on experimental trends of the last 5+ years
\cite{figgatt,mandviwalla,Zhang3,Zhang4,pokharel,thorvaldson,abughanem} and progress in circuit efficiency \cite{yong,dasilva,dreier,nie,silva2}, it
does not appear likely that this number $N$ will dramatically increase in the next several years without a significant technological breakthrough (extending $N$ into the $10$s or $100$s of qubits). By contrast, current hardware (especially superconducting qubits which support parallel gates) is already powerful enough to implement cost oracle operations up to hundreds of qubits, making diffusion the technological bottleneck limiting QAA. Therefore, another avenue for future research is exploring QAA performance using less than full $N$-qubit diffusion, instead replaced by multiple low qubit-sized diffusions in parallel.  This would significantly reduce circuit depth, but also introduces more $\theta$ degrees of freedom which complicate the algorithm.    If small diffusions in parallel can produce probabilities comparable to full $N$-qubit diffusion, then QAA could potentially be a viable near term quantum algorithm alongside QAOA for solving combinatorial optimization problems.

\section*{Acknowledgments}

Any opinions, findings, conclusions or recommendations expressed in this
material are those of the author(s) and do not necessarily reflect the views of
AFRL. This project was supported in part by an appointment to the NRC Research
Associateship Program at AFRL, administered by the Fellowships Office of the
National Academies of Sciences, Engineering, and Medicine.

\section*{Data \& Code Availability}

The data and code files that support the findings of this study are available
from the corresponding author upon reasonable request.

\bibliographystyle{unsrt}
\bibliography{main}

\appendix

\section{Cost Functions \& Distributions }\label{sec:appendixA}

\subsection{C$(Z)$ Weights} \label{sec:prob_instance}

Found in sections III \& IV are references to a particular problem cases of
linear cost functions which can be found here. Given in equation \ref{Aeqn.simple_cz} is $\mathbb{W}_1$, the set of positive integers from $1$ to $N$, which are the sets of weights for figures \ref{Fig.Q10_ssp_cp}, \ref{Fig.ssp_prob_plots}, and \ref{Fig.Q40_prob_plots}, as well as experiments 1 and 2 in section IV.  Also shown in figure \ref{Fig.Q40_prob_plots} is a second $40$-qubit cost function corresponding to the weights $\mathbb{W}_3$ given in equation \ref{Aeqn.random_cz}.

\begin{eqnarray} \mathbb{W}_{1} &=& \{ 1,2,3,...,N\}\label{Aeqn.simple_cz}
 \\ \mathbb{W}_{2} &=& \{ -44, -35, -33, -32, -23, -20, -11, -11, \nonumber \\ & &  -10, -4, 2, 6, 9, 11, 11, 17, 21, 34, 40, 43 \} \label{Aeqn.Q20_random_cz}
 \\
 \mathbb{W}_{3} &=& \{ -731,-722,-676,-668,-663,-662,-564,
\nonumber \\ & &-563,-555,-409,-209,-189,-135,-43, \nonumber \\ &
&1,3,28,48,73,127,139,156,160,286,307, \nonumber \\ & & 308,427,
461,490,512,548,551,568, \nonumber \\ & & 583,589,642,776,917,929,948,949
\}\label{Aeqn.random_cz} \end{eqnarray}

\subsection{Experiment Collective States} 
\label{sec:exp_states}

\begin{figure}[ht] \centering \includegraphics[scale=.50]{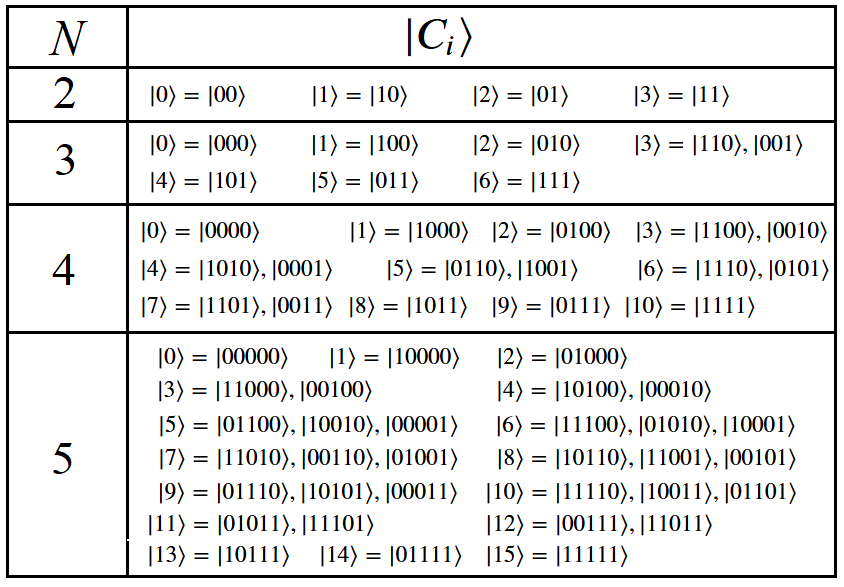} \caption{ A table displaying all of the collective states
$|C_i\rangle$ for experiments 1 and 2, corresponding to a linear C$(Z)$ composed of the weights $\mathbb{W}_1$ from equation \ref{Aeqn.simple_cz} up to $N=5$, as well as the basis states
$|Z_i\rangle$ contained within them. } \label{AFig.Simple_SSP_Table}
\end{figure}

For the results of experiments 1 and 2 found in section IV, which use $\Uc(\ps)$ encoding linear C$(Z)$ according to equation \ref{Aeqn.simple_cz}, figure \ref{AFig.Simple_SSP_Table} below provides a table all of the
collective states $|C_i\rangle$ along with the basis states $|Z_i\rangle$
contained within them.

\section{Fidelity Metric} \label{sec:f_metric}

Here we discuss in full mathematical detail the calculation of $f$, the metric of performance used in section IV. We begin with equation \ref{Eqn.Meas} for Meas.($|Z_i\rangle$), which is simply the experimentally observed probability of a basis state. `Shots' refers to the number of times a quantum circuit was run, producing that number of measurement results, which in this study is $10,000$. Given in equation \ref{Eqn.Delta_P_metric1} is the the quantity $\Delta
P_{i}$, which is the difference in probability between measured counts of a
$|Z_i\rangle$ basis state (equation \ref{Eqn.Meas}) and its theoretical expected probability given by equation \ref{Eqn.first_k_non} or \ref{Eqn.first_k_binary}.

\begin{eqnarray} \textrm{Meas}.(|Z_i\rangle)\label{Eqn.Meas} &=&
\frac{\textrm{Counts}(|Z_i\rangle)}{\textrm{shots}} \\ \Delta P_{i} &=&
\big{|}\hspace{0.03cm} \textrm{Meas}.(|Z_i\rangle) - | \langle Z_i | \Psi
\rangle |^2 \hspace{0.03cm} \big{|}\label{Eqn.Delta_P_metric1}
\end{eqnarray}

The problem with
using raw probability difference to evaluate performance is that good or bad is relative for each basis state. This was the motivation for introducing $f$, which compares each $\Delta P_i$ against probabilities corresponding to a completely decohered equal superposition state, given below in equation \ref{Eqn.Delta_P_noise} as $\Delta \tilde{P}_{i}$.
 
\begin{eqnarray} \Delta \tilde{P}_{i} &=& \big{|}\hspace{0.03cm}
\frac{1}{2^N} - | \langle Z_i | \Psi \rangle |^2 \hspace{0.03cm}
\big{|}\label{Eqn.Delta_P_noise}       
\end{eqnarray}

The computation of $f$ is based on the ratio between root mean square calculations of $\Delta P_{i}$ and $\Delta \tilde{P}_{i}$ values from all 100 circuits, as shown in equation \ref{Eqn.RMS_i} (note that $i$ refers to the basis state while $j$ refers to one of the 100 quantum circuits).   For each basis state $|Z_i\rangle$ we compute $f_i$ according to equation \ref{Eqn.f_i} (where
$\widetilde{\textrm{RMS}}$ is obtained from using $\Delta \tilde{P}_{i}$ values in equation \ref{Eqn.RMS_i}).  Averaging all $2^N$ $f_i$ together yields $f_{\textrm{exp}}$, given in equation \ref{Eqn.f_exp}.

\begin{eqnarray} \textrm{RMS}_i &=& \sqrt{ \frac{1}{100} \sum_j^{100} \Delta
P_{ij}^2 } \label{Eqn.RMS_i} \\ f_{i} &=& 1 -
\frac{\textrm{RMS}_i}{\widetilde{\textrm{RMS}}_{i}} \label{Eqn.f_i} \\
f_{\textrm{exp}} &=& \frac{1}{2^N} \sum_i^{2^N} f_i \label{Eqn.f_exp}
\end{eqnarray}

\section{Symmetry Proofs for Linear C$(Z)$}\label{sec:appendixB}

To assist with the forthcoming derivations, given below in figure \ref{AFig.Q10_Histo} is a histogram of all $C_i$ solutions to the 10-qubit linear C$(Z)$ composed of the set of weights $\mathbb{W}_1$ from equation \ref{Aeqn.simple_cz}.  Each black circle shown in the figure corresponds to one collective state $|C_i\rangle$ containing $N_i$ basis states.  

\begin{figure}[ht] \centering \includegraphics[scale=.4]{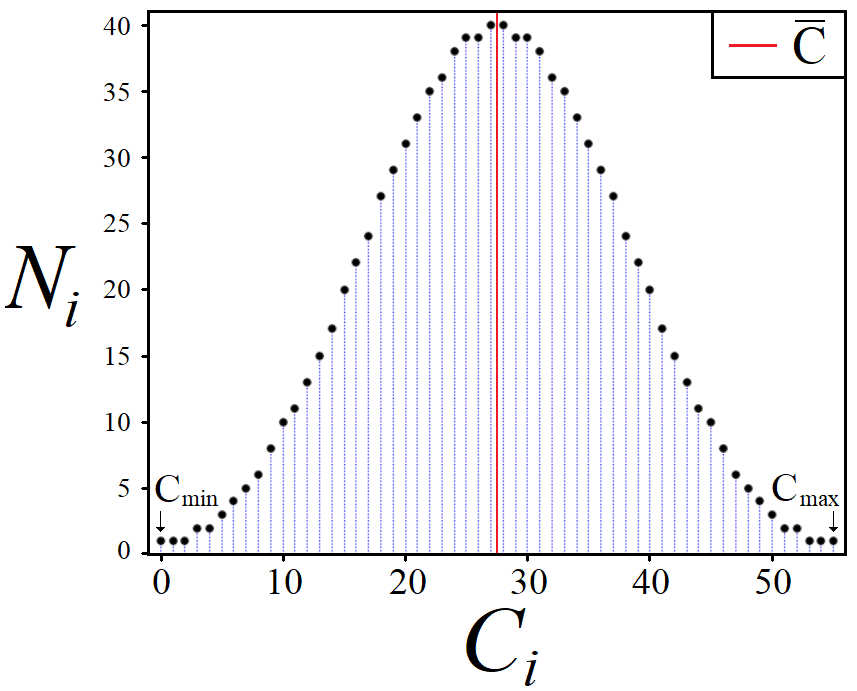} \caption{ Histogram of all $C_i$ solutions to a 10-qubit C$(Z)$ using the weights $\mathbb{W}$ given in equation \ref{Aeqn.simple_cz}. (red line) The value $\bar{C}$ from equation \ref{Eqn.c_bar1}. } \label{AFig.Q10_Histo}
\end{figure}

Equation \ref{Eqn.abar_cbar} from section III states that for any $\Uc(\ps)$ encoding a linear C$(Z)$ the resulting phase of $\bar{\alpha}$ is equal to the $\bar{C} \cdot \ps$.  Here we shall derive the equations leading to this relation, showing explicitly the computation of $\bar{\alpha}$ after the first two applications of $\Uc(\ps)$. The results which follow are specifically for C$(Z)$
according to equation \ref{Eqn.ssp_cost_function2} and do not generalize to
more complex problems such as QUBO. We begin with equation \ref{Aeqn.inverse_z1} below which defines a pair of inverse solutions $Z_i$ and $\neg Z_i$, two binary strings that have equal and opposite $0$ and $1$ values on every bit such that their binary sum equals $2^N-1$ (base-10).  For linear C$(Z)$ the evaluation of any pair of inverse $Z_i$ is equal to the full sum of $\mathbb{W}$.

\begin{eqnarray} Z_i + \neg Z_i &=& 111...1 = 2^N-1 \label{Aeqn.inverse_z1} \\
\textrm{C}(Z_i) + \textrm{C}(\neg Z_i) &=& \sum_i^N W_i = W{_\textrm{sum}}\label{Aeqn.inverse_z2} \end{eqnarray}

Using equation \ref{Aeqn.inverse_z2} we can now prove equation \ref{Eqn.abar2}, the mean cost
function value $\bar{C}$ according to equation \ref{Eqn.c_bar1} is always
equal to half of the sum of any pair of inverse solutions, given by equations
\ref{Aeqn.cbar_proof1} - \ref{Aeqn.cbar_proof4} below.
\begin{eqnarray} \bar{C}  &=& \frac{1}{2^N} \sum^{2^N}_{i} \textrm{C}(Z_i)
\label{Aeqn.cbar_proof1} \\  &=& \frac{1}{2^N} \sum^{2^{N-1}}_{i}
\textrm{C}(Z_i) + \textrm{C}(\neg Z_i) \label{Aeqn.cbar_proof2} \\  &=& \frac{2^{N-1} \cdot W_{\textrm{sum}} }{2}    \label{Aeqn.cbar_proof3} \\ &=& \frac{ \textrm{C}(Z_j) + \textrm{C}(\neg Z_j)
}{2}    \label{Aeqn.cbar_proof4} \end{eqnarray}

For the next derivation let us define $\Delta C_i$, given in equation \ref{Aeqn.delta_c1}, which is simply the difference between the mean cost function value and any $C(Z_i)$.
\begin{eqnarray} \Delta C_i = \bar{C} - \textrm{C}(Z_i)   \label{Aeqn.delta_c1} \end{eqnarray}

Illustrated in figure \ref{AFig.Q10_Histo} is the symmetry property of linear C$(Z)$ that all solutions come in equal and opposite pairs that are equidistant from $\bar{C}$.  Given in equations \ref{Aeqn.delta_c2} - \ref{Aeqn.delta_c4} below is the proof that these symmetric solutions are in fact pairs of inverse $Z_i$.
\begin{eqnarray} 
\textrm{C}(Z_i) + \textrm{C}(\neg Z_i) &=& 2 \bar{C}
\label{Aeqn.delta_c2} \\ \bar{C} - \Delta C_i + \textrm{C}(\neg Z_i) &=&
2\bar{C}  \label{Aeqn.delta_c3} \\  \textrm{C}(\neg Z_i) &=& \bar{C} +
\Delta C_i   \label{Aeqn.delta_c4} \end{eqnarray}

Using equations \ref{Aeqn.delta_c1} and \ref{Aeqn.delta_c4}, we can now
prove the relation between $\bar{C}$ and $\bar{\alpha}$ given in equation
\ref{Eqn.abar_cbar} of section III.  The following equations use the
amplitudes $\alpha_i$ corresponding to the state $|\Psi\rangle =
\Uc(\ps)|s\rangle$, the first oracle application acting on the equal superposition state. By
rewriting the summation for $\bar{\alpha}$ first in terms of inverse $Z_i$
pairs and then substituting in $\Delta C_i$ and $\bar{C}$ equivalents, we
arrive at equation \ref{Aeqn.abar_proof6}. 

\begin{align}
	\bar{\alpha}_1 &= \frac{1}{2^N} \sum_{j}^{2^N} \alpha_j \label{Aeqn.abar_proof1} \\
	&= \frac{1}{2^N \sqrt{2^N}} \sum_{j}^{2^N} e^{i C(Z_j)\cdot \ps} \label{Aeqn.abar_proof2} \\
	&= \frac{1}{(2^N)^{3/2}} \sum_{j}^{2^{N-1}} \left( e^{i C(Z_j)\cdot \ps} + e^{i C(\neg Z_j)\cdot \ps} \right) \label{Aeqn.abar_proof3} \\
	&= \frac{1}{(2^N)^{3/2}} \sum_{j}^{2^{N-1}} e^{i \bar{C}\cdot \ps}
	\left( e^{-i \Delta C_j \cdot \ps} + e^{i \Delta C_j \cdot \ps} \right) \label{Aeqn.abar_proof4} \\
	&= \frac{e^{i \bar{C}\cdot \ps}}{(2^N)^{3/2}}
	\sum_{j}^{2^{N-1}} 2\cos(\Delta C_j \cdot \ps) \label{Aeqn.abar_proof6}
\end{align}

To summarize, equation \label{Aeqn.abar_proof6} shows that the value of $\bar{\alpha}_1$ after the first application of $\Uc$ is equal to $e^{i \bar{C}\cdot p_{\textrm{s}}}$ times a constant, confirming the relation given by equation
\ref{Eqn.abar_cbar}, which in turn is used to derive equation \ref{Eqn.ssp_ps_opt1} for finding $\ps$ values which create a $\pi$ phase difference between $\bar{\alpha}$ and any $|C_i\rangle$. Importantly, \ref{Aeqn.abar_proof6} shows that the phase of $\bar{\alpha}$ is proportional to $\bar{C}$, but the full computation of $\bar{\alpha}$ still requires complete knowledge of all $C_i$. 

So far we have shown that equation \ref{Eqn.abar_cbar} holds for the first oracle application, so next we shall prove that it holds for the second as well. Given in equation \ref{Aeqn.alpha_j_1} below are the
amplitudes $\alpha'_j$ corresponding to the state  $|\Psi\rangle=\Us(\pi)\Uc(\ps)|s\rangle$.

\begin{eqnarray} \alpha'_j = \frac{e^{i \textrm{C}(Z_j)\cdot
\ps}}{\sqrt{2^N}} - 2 |\bar{\alpha}_1|e^{i \bar{C}\cdot
\ps} \label{Aeqn.alpha_j_1} \end{eqnarray}

Equations
\ref{Aeqn.abar2_proof1} - \ref{Aeqn.abar2_proof4} below show the calculation
of $\bar{\alpha}_2$, the mean amplitude following the second oracle
application.

\begin{eqnarray} \bar{\alpha}_2  &=& \frac{1}{2^N} \sum^{2^N}_{j}
\frac{e^{i2 \textrm{C}(Z_j)\cdot \ps}}{\sqrt{2^N}} - 2
|\bar{\alpha}_1|e^{i ( \bar{C} + \textrm{C}(Z_j)) \cdot \ps}
\label{Aeqn.abar2_proof1} \\ &=& \frac{1}{2^N} \sum_{j}^{2^{N-1}}
\frac{\Big( e^{i 2 \textrm{C}(Z_j) \cdot \ps} + e^{i 2
\textrm{C}(\neg Z_j) \cdot \ps} \Big)}{\sqrt{2^N}} \nonumber  \\
& & \hspace{0.1cm} -  2|\bar{\alpha}_1| e^{i \bar{C}\cdot \ps}
\Big( e^{i  \textrm{C}(Z_j) \cdot \ps} + e^{i
\textrm{C}(\neg Z_j) \cdot \ps} \Big) \label{Aeqn.abar2_proof2}
\\ &=& \frac{e^{i 2 \bar{C}\cdot \ps}}{2^N} \sum_{j}^{2^{N-1}}
\frac{ \Big( e^{-i 2 \Delta C_j \cdot \ps} + e^{i 2 \Delta C_j
\cdot \ps} \Big) }{\sqrt{2^N}} \nonumber  \\ & & \hspace{0.9cm} -
2|\bar{\alpha}_1| \Big( e^{-i  \Delta C_j \cdot \ps} + e^{i
\Delta C_j \cdot \ps} \Big) \label{Aeqn.abar2_proof3} \\ &=&
\frac{ e^{i 2 \bar{C}\cdot \ps}}{2^{N-1}}
\sum_{j}^{2^{N-1}}\frac{\textrm{cos}(2 \Delta C_j \cdot
\ps)}{\sqrt{2^N}} \nonumber \\ && \hspace{1.3cm} -
2|\bar{\alpha}_1| \textrm{cos}(\Delta C_j \cdot \ps)
\label{Aeqn.abar2_proof4} \end{eqnarray} 

In equation \ref{Aeqn.abar2_proof4} above we see that the phase of
$\bar{\alpha}_2$ is equal to $2 \bar{C} \cdot \ps$, but once again the full value of $\bar{\alpha}_2$ requires complete knowledge of every amplitude value. To conclude, we note that the derivation of two iterations given above is general, showing that phase relation given in equation \ref{Eqn.abar_cbar} holds, also supported by the simulation results of section III, but does not constitute a complete proof by induction.

\section{Resonance Width in Grover's}\label{sec:appendixD}

The QAA algorithm using iterations of $\Us(\theta)
\Ug(\phi)$ in Alg.~\ref{Alg.AmpAmp} is optimal when both free
parameters are $[\theta,\phi] = [\pi,\pi]$ \cite{grover,boyer,zalka}. However,
it is important to understand the behavior of the algorithm when these phases
differ, particularly when considering noisy gates which lead to imperfect
implementations of the oracle \cite{shenvi, long, long2}.  In this appendix we
show the relation between peak achievable probabilities as a function of
$\phi$ around $\pi$ following closely the approach of \cite{hsieh}. The
resonance phenomena demonstrated here for Grover's at $\phi=\pi$ is the same
as those shown in figure \ref{Fig.solving_for_T} for each of the individual $|C_i\rangle$ states \cite{koch1,bench,koch2}. Equation \ref{Eqn.ssp_ps_opt3} from section III produces near optimal $\ps$ values for linear C$(Z)$, but in general determining $\ps$ values which align with $|C_i\rangle$ resonance peaks for QUBO and harder problem instances is an open research question \cite{zhukov}.  We begin with equation \ref{AEqn.s_columnvec} below, the equal superposition
state $|s\rangle$ expressed as a column vector.  For simplicity the initial
amplitudes of $|m\rangle$ and $|n\rangle$ are $\sin \beta = \sqrt{N_m/2^N}$
and $\cos \beta = \sqrt{N_n/2^N}$ respectively.
\begin{equation} \ket{s}=\sin{\beta}\ket{m}+\cos{\beta}\ket{n} =
\begin{bmatrix} \sin \beta \\ \cos \beta \end{bmatrix}.
\label{AEqn.s_columnvec} 
\end{equation}

Next we need the 2x2 matrix form of $\Us(\theta)
\Ug(\phi)$, which we call $G$ in equation \ref{AEqn.G_mat1}
below.
\begin{eqnarray} G &\equiv& \Us(\theta)  \Ug(\phi)
\label{AEqn.G_mat1} \\  &=& \left[ \begin{smallmatrix} e^{i \phi}\bigl(1+(e^{i
\theta}-1) \sin ^2(\beta)\bigr) & \left(e^{i \theta}-1\right) \sin (\beta) \cos
(\beta) \\ e^{i \phi}(e^{i \theta}-1) \sin (\beta) \cos (\beta) & 1+\left(e^{i
\theta}-1\right) \cos ^2(\beta) \end{smallmatrix}  \right]   \label{AEqn.G_mat2}
\end{eqnarray}

Since $G$ given in equation \ref{AEqn.G_mat2} is a unitary matrix, it can be
represented generically by the matrix decomposition given in equation
\ref{Aeqn.decomp}, where $U$ is some unitary 2x2 matrix and $D$ is a diagonal
matrix with values $\lambda_+$ and $\lambda_-$. The matrix $U$ has columns given
by the normalized eigenvectors of equation \ref{Aeqn.eigenvecs}, with
eigenvalues given in equations \ref{Aeqn.eigenvalues} and \ref{Aeqn.cos_w}.
\begin{eqnarray} G &=& U^\dagger DU,\label{Aeqn.decomp} \\[6pt] u_+ &=&
\begin{bmatrix} e^{-i\frac\phi2}\cos{x}\\ \sin{x} \end{bmatrix}\quad
u_-=\begin{bmatrix} -\sin{x}\\ e^{i\frac\phi2}\cos{x} \end{bmatrix}
\label{Aeqn.eigenvecs} \\[6pt] \lambda_{\pm} &=&
-e^{i\frac{(\theta+\phi)}{2}\pm w} \label{Aeqn.eigenvalues} \\[6pt] \cos w
&=& \cos \Bigl(\frac{\phi-\theta}{2}\Bigr)-2 \sin\frac{\phi}{2} \sin
\frac{\theta}{2} \sin ^2\beta. \label{Aeqn.cos_w} \end{eqnarray}

Using equations \ref{Aeqn.decomp} - \ref{Aeqn.cos_w}, we can write the general
form of the Grover operator $G$ after $t$ iterations, given in equation
\ref{Aeqn.Gt}.
\begin{equation} G^t= \left[\begin{smallmatrix} e^{iwt} \cos ^2x+e^{-iwt} \sin
^2x & i e^{-i \frac{\phi}{2}} \sin (wt) \sin (2 x) \\ ie^{i \frac{\phi}{2}} \sin
(wt) \sin (2 x) & e^{iwt} \sin ^2x+e^{-iwt} \cos ^2x \end{smallmatrix}\right]
\label{Aeqn.Gt} \end{equation}

Next we need expressions for the angle $x$, which can be found by using the
equation $G u_+ = \lambda_+ u_+$, or equivalently by setting the off-diagonal
elements of $U^\dagger G U$ to 0. Solving this first expression yields the
equations for $\sin x$ and $\cos x$ given in equations \ref{Aeqn.sinx} -
\ref{Aeqn.lm} below.
\begin{eqnarray} \sin x&=& l_m ^{-\frac12}\sin \frac{\theta}{2} \sin (2 \beta)
\label{Aeqn.sinx} \\ \cos x&=& l_m ^{-\frac12}\Bigl[\sin w+\sin
\Bigl(\frac{\phi-\theta}{2}\Bigr) \label{Aeqn.cosx}  \\ && \hspace{0.5cm} +2
\cos \frac{\phi}{2} \sin\frac{\theta}{2} \sin ^2\beta\Bigr]  \nonumber  \\[4pt]
l_m &=&\big(\sin \frac{\theta}{2} \sin (2 \beta)\big)^2 + \Bigl[\sin w +
\sin \Bigl(\frac{\phi-\theta}{2}\Bigr) \label{Aeqn.lm}  \\ &&
\hspace{0.5cm} +2 \cos \frac{\phi}{2} \sin\frac{\theta}{2} \sin
^2\beta\Bigr]^2  \nonumber \end{eqnarray}

Using the equations above for $G^t$, we can compute the amplitude of $|m\rangle$
at any iteration $t$ using equation \ref{Aeqn.m_Gt_s} below. 
\begin{eqnarray} \braket{m|G^t|s} &=&    \sin\beta\cos (wt) \label{Aeqn.m_Gt_s}
\\ &+& i\big( e^{i\phi/2}\cos\beta\sin(2x) +\sin\beta\cos(2x)\big) \sin (wt)
\nonumber \end{eqnarray}

Taking the absolute value squared of equation \ref{Aeqn.m_Gt_s} yields $P_m(t) =
|\! \braket{m|G^t|s} \! |^2$, the probability of measuring $\ket m$ after $t$
iterations. In the large $2^N$ limit, this probability simplifies significantly
to equation \ref{Aeqn.Pt_approx} below.
\begin{equation} P_m(t)\approx \sin^2(2x) \sin^2(wt) \label{Aeqn.Pt_approx}
\end{equation}

The probability $P_m(t)$ is maximal at $t= \frac\pi{2w}$, which after
substituting into the equation above yields equation \ref{Aeqn.Pmax1} for
$P_\textrm{max}$, the maximum achievable probability for $|m\rangle$ as a
function of $\phi$ and $\theta$.
\begin{equation} P_{\max} \approx \frac{4\sin^2\frac{\theta}{2}} {2^N\sin^2 \!
\big( \frac{\theta-\phi}{2}\big) +4\sin\frac{\theta}{2}\sin\frac{\phi}{2}\cos \!
\big( \frac{\theta-\phi}{2} \big)}\label{Aeqn.Pmax1} \end{equation}

\begin{figure}[ht] \centering \includegraphics[scale=0.3]{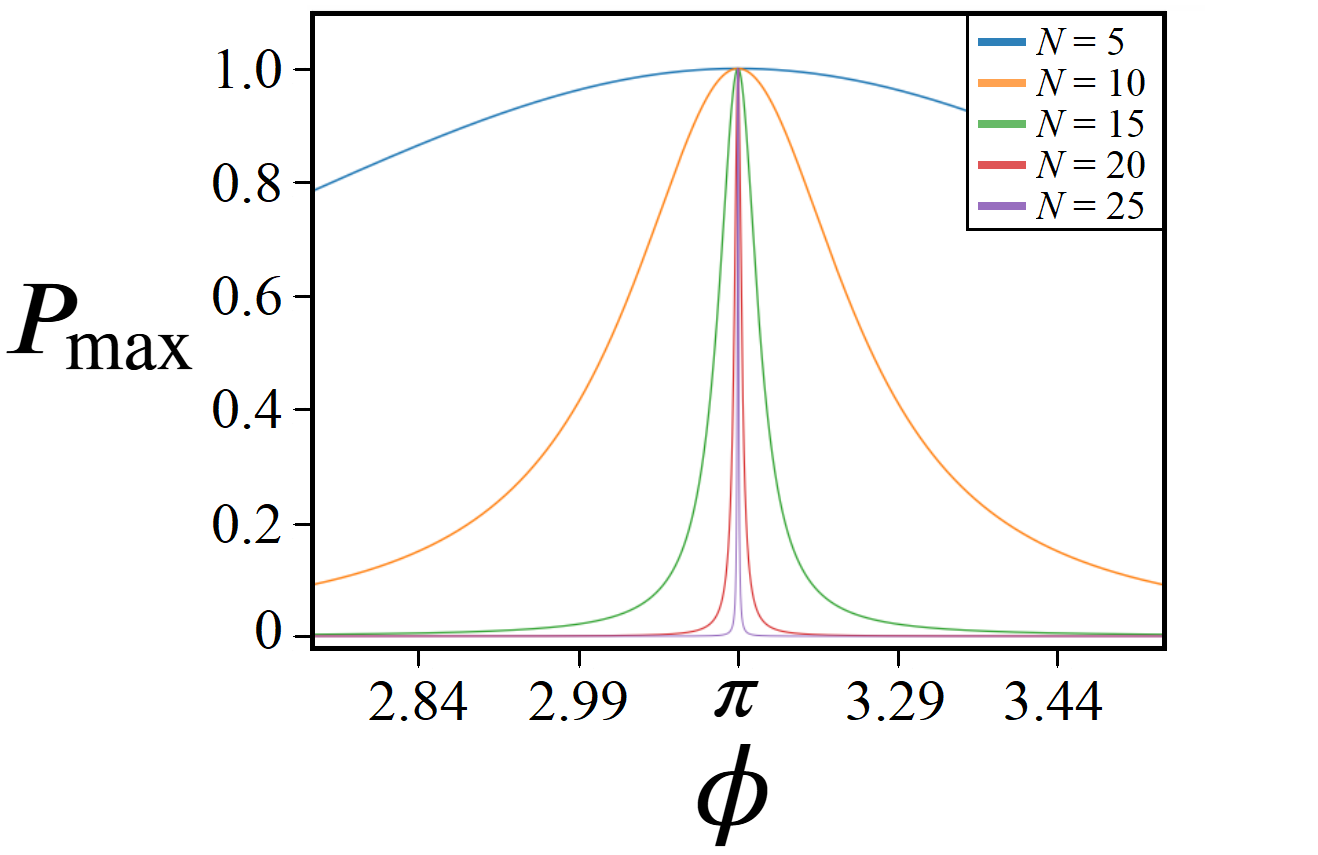}
\caption{ Peak achievable probability $P_{\textrm{max}}$ from equation
\ref{Aeqn.Pmax2} as a function of oracle phase $\phi$ for various $N$-qubit
problem sizes. Each plot shows the resonance behavior around $\phi=\pi$ for
Grover's algorithm using $\Us(\pi)$ for diffusion searching for a
single marked state $N_m=1$.} \label{Fig.resonance} \end{figure}

And finally, we are interested in the case of $\theta=\pi$, yielding equation
\ref{Aeqn.Pmax2} for $P_{\textrm{max}}$ as a function of $\Ug$ oracle
angle $\phi$. Shown in figure \ref{Fig.resonance} are five plots for various
problem sizes $N$, illustrating the resonance behavior sharply peaked around
$\phi = \pi$, becoming narrower with increasing problem size $N$.
\begin{equation} P_{\max{}} =
\frac{1}{\sin^2\frac{\phi}{2}+\frac{2^N}{4}\cos^2\frac{\phi}{2}}.
\label{Aeqn.Pmax2} \end{equation}

Figure \ref{Fig.resonance} illustrates the degree of precision in $\phi$
necessary for Grover's algorithm as problem size $N$ increases, also
representing the required accuracy in $\ps$ for cost oracles
\cite{bench,koch2}.  We can quantify this required precision with equation
\ref{Aeqn.fwhm} below, the full width half maximum (FWHM) for problem sizes
$N>2$
\begin{equation} \delta\phi = 2\cos^{-1} \Big(\frac2{\sqrt{2^N-4}}
\Big)\label{Aeqn.fwhm} \end{equation}

\section{Quantum Circuits}
\label{sec:appendixE}

\begin{figure}[H] \centering 
\input{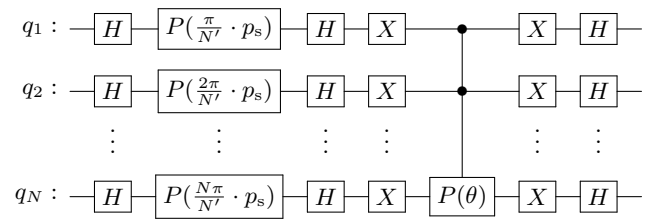}
\caption{ Quantum circuit for experiments 1 and 2} \label{Fig.Exp1_2_qc} \end{figure}

Here we present the quantum circuits for experiments 1-3 from section IV, with further details on the decomposition of the C$^N$-$P(\theta)$ gate given in the next subsection. Shown in figure \ref{Fig.Exp1_2_qc} below is the quantum circuit for experiments 1 and 2, corresponding to the state $\Us(\theta) \Uc(\ps)|s\rangle$ given in equation \ref{Eqn.first_k_non}. The oracle $\Uc$ here encodes $\mathbb{W}_1$ from equation \ref{Aeqn.simple_cz}.  Experiment 1 uses $\theta=\pi$ while varying $\ps$, while conversely experiment 2 uses $\ps=1$ while varying $\theta$. The value of $N'$ is given by equation \ref{Eqn.N_prime} below.
\begin{eqnarray} N' = \sum_{i=1}^{N} i = \frac{1}{2}( N^2 + N )
\label{Eqn.N_prime} \end{eqnarray}

\begin{figure*}[t]
\begin{tabular}{|c|c|cc|cc|cc|cc|} \hline \multicolumn{2}{|c}{} &
\multicolumn{2}{|c|}{Fez} & \multicolumn{2}{|c|}{Torino} &
\multicolumn{2}{|c|}{Kyiv} & \multicolumn{2}{|c|}{Brisbane}\\ \hline Experiment
& Qubits  &  Depth  &  2q Gates  &  Depth  &  2q Gates  &  Depth  &  2q Gates  &
Depth  &  2q Gates  \\ \hline \multirow{4}{4em}{$p_s$} &    2     &   23    &
2      &   23    &     2      &   28    &     2      &   28    &     2      \\ &
3     &   51    &     9      &   57    &     9      &   71    &     9      &
56    &     9      \\ &    4     &   113   &     20     &   113   &     20     &
111   &     20     &   124   &     20     \\ &    5     &   239   &     48     &
239   &     48     &   235   &     48     &   259   &     48     \\ \hline
\hline \multirow{4}{4em}{Diffusion} &    2     &   23    &     2      &   23
&     2      &   28    &     2      &   28    &     2      \\ &    3     &   51
&     9      &   51    &     9      &   60    &     9      &   61    &     9
\\ &    4     &   113   &     20     &   119   &     20     &   111   &     20
&   130   &     20     \\ &    5     &   239   &     48     &   239   &     48
&   228   &     48     &   296   &     63     \\ \hline \hline
\multirow{4}{4em}{Grover's} &    2     &   32    &     4      &   32    &
4      &   42    &     4      &   27    &     4      \\ &    3     &   95
&     21     &   95    &     21     &   124   &     21     &   127   &     21
\\ &    4     &   219   &     43     &   219   &     43     &   232   &
43     &   246   &     43     \\ &    5     &   493   &    126     &   534
&    126     &   626   &    126     &   615   &    126     \\ \hline
\end{tabular} \caption{Circuit depth and number of 2 qubit gates across all
experiments and devices.} \label{Fig.gate_table} \end{figure*}

\begin{figure}[H] \centering 
\input{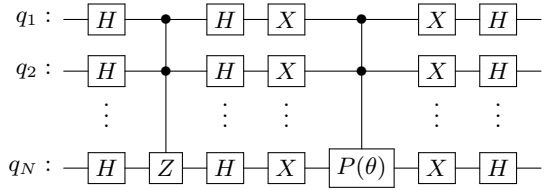}
\caption{ Quantum circuit for experiment 3} \label{Fig.Exp3_qc} \end{figure}

The quantum circuit for experiment 3 is given in figure \ref{Fig.Exp3_qc}, using $\Ug(\pi)$ as the oracle while varying $\theta$ in the diffusion operator. The C$^N$-Z operator shown in the figure was implemented using C$^N$-$P(\pi)$ as discussed in the next subsection. The quantum state produced from this circuit is equation \ref{Eqn.first_k_binary}.  Given in figure
\ref{Fig.gate_table} is a table detailing the circuit depth and 2-qubit gate
count for all three experiments on IBMQ's four processors, where circuit depth is
defined as parallel gate layers \cite{mckay}.

\subsection{Circuit Construction of
\ensuremath{C^N-P(\theta)}} 

The most resource-intensive component of both the Grover oracle and the
diffusion operator is the fully-entangling C$^N$-P$(\theta)$ gate. This
multiqubit gate must be transpiled into 1 \& 2-qubit instructions, for which there are several approaches. In this work, we opt for a simple decomposition that is not
optimized for any particular architecture. We send the same quantum circuit to
both IBMQ and IonQ, decomposing $C^NP(\theta)$ into 2-qubit CX and single
qubit $P(\theta)$ operations. For IBMQ we transpiled directly to the backend's qubits, while for IonQ qubit
selection and gate transpilation were left to the hardware provider.

The decomposition procedure used in this study is adapted from
\cite{barenco, schuchthesis}. Consider a sequence of binary variables $z_i$ for
$i\in [1, N]$. We want to apply a $2\times2$ unitary gate $U$, in our case the phase gate $P(\theta)$ from equation \ref{Eqn.phase_gate}, to the $N^\mathrm{th}$ qubit if and only if $z_i=1$ for all
$i\in [1, N-1]$. This operation can be expressed by the identity given in equation \ref{Aeqn.grey}.
\begin{align} 2^{m-1}(z_1 \wedge z_2 \wedge \cdots \wedge z_m) =\sum_{k_1}
z_{k_1}- \nonumber \\\sum_{k_1<k_2}(z_{k_1} \oplus
z_{k_2})+\cdots+(-1)^{m-1}(z_1 \oplus \cdots \oplus z_m) \label{Aeqn.grey} \end{align}

The operation above can be implemented using a Gray sequence to determine the control and
target qubits. An $N$-bit Gray code is the sequence of $2^N$ binary numbers with
adjacent elements in the sequence differing by a single bit flip. The particular
Gray sequence that we use can be determined recursively as follows. We denote a
Gray sequence by $G_N = (g_1, g_2,\dots,g_{2^N})$, with the base case Gray sequence given by $G_1 = (0, 1)$. We can recursively define a Gray sequence given by
$G_{N+1} = (0G_N, 1 \tilde G_N)$, where $\tilde G$ is the reverse ordering of
$G$. For example, this produces $G_2 = (00, 01, 11, 10)$ followed by $G_3 = (000, 001,
011, 010, 110, 111, 101, 100)$. To implement this code as gates, we note that
the leading $1$ corresponds to the target qubit and the remaining $1$'s
correspond to the XOR combination for which we apply the gate $V$ or
$V^\dagger$. The parity of the bits determine which is applied: $V$ if even or
$V^\dagger$ if odd. Given in figure \ref{Aeqn.gray_circuit} below is the quantum circuit used in this study for implementing a C$^N$-$P(\theta)$ gate, with further details given in figure \ref{Aeqn.gray_unitaries}.

\begin{figure}[ht]
\[ \Qcircuit @C=0.4em @R=0.2em @!R { \lstick{q_{1}  }  & \qw & \gate{U_1} & \qw
& \multigate{1}{U_2} & \qw & \multigate{2}{U_3}  & \qw & & & & &  \qw &
\multigate{4}{U_n} & \\ \lstick{q_{2}  }  & \qw & \qw      & \qw & \ghost{U_2}
& \qw & \ghost{U_3}  & \qw & & & & &  \qw & \ghost{U_n} & \\ \lstick{q_{3}  }  &
\qw & \qw      & \qw & \qw                  & \qw & \ghost{U_3} & \qw & & \cdots
& & & \qw & \ghost{U_n}\\ \nghost{q_{2}   }   & & & \vdots & & & & & & & & & &
\nghost{U_n}& \\ \lstick{q_{n}  }  & \qw & \qw      & \qw & \qw
& \qw & \qw & \qw & & & & & \qw & \ghost{U_n} &} \]
\caption{Quantum circuit for implementing the Gray code.} \label{Aeqn.gray_circuit}
\end{figure}
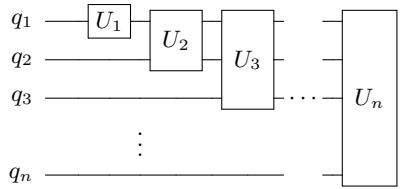

\begin{figure*}[t] \[ \vcenter{\Qcircuit @C=0.4em @R=0.2em @!R { \nghost{q_{1}
} & \lstick{q_{1}  } & \gate{U_1} & \qw & }} = \vcenter{\Qcircuit @C=0.4em
@R=0.2em @!R { \nghost{q_{1}  } & \lstick{q_{1}  } & \gate{V} & \qw & }} = \vcenter{\Qcircuit @C=0.4em
@R=0.2em @!R { \nghost{q_{1}  } & \lstick{q_{1}  } & \gate{P(\theta')} & \qw & }} \qquad
\vcenter{\Qcircuit @C=0.4em @R=0.2em @!R { \nghost{{q}_{1}  } & \lstick{{q}_{1}
} & \multigate{1}{U_2} & \qw &\\ \nghost{{q}_{2}  } & \lstick{{q}_{2}  } &
\ghost{U_1}   & \qw &}} = \vcenter{\Qcircuit @C=0.4em @R=0.2em @!R {
\nghost{q_{1}  }& \lstick{q_{1}  } & \ctrl{1} & \qw & \ctrl{1} & \qw & \qw & \\
\nghost{q_{1}  }& \lstick{q_{2}  } & \targ & \gate{V^\dagger} & \targ & \gate{V}
& \qw &}} \] \[ \vcenter{\Qcircuit @C=0.4em @R=0.2em @!R { \nghost{{q}_{1}  } &
\lstick{{q}_{1}  } & \multigate{2}{U_3} & \qw &\\ \nghost{{q}_{2}  } &
\lstick{{q}_{2}  } & \ghost{U_1}   & \qw & \\ \nghost{{q}_{3}  } &
\lstick{{q}_{3}  } & \ghost{U_1} & \qw &}} = \vcenter{\Qcircuit @C=0.4em
@R=0.2em @!R { \nghost{{q}_{1}  }& \lstick{q_{1}  } &\ctrl{2} & \qw & \qw & \qw
& \ctrl{2} & \qw & \qw & \qw & \qw &\\ \nghost{{q}_{2}  }& \lstick{q_{2}  } &
\qw & \qw & \ctrl{1} & \qw & \qw & \qw & \ctrl{1} & \qw & \qw  &\\
\nghost{{q}_{3}  }& \lstick{q_{3}  } & \targ & \gate{V^\dagger} & \targ &
\gate{V} & \targ & \gate{V^\dagger} & \targ & \gate{V} & \qw & }} \] \[
\vcenter{\Qcircuit @C=0.4em @R=0.2em @!R { \nghost{{q}_{1}  } & \lstick{{q}_{1}
} & \multigate{3}{U_4} & \qw &\\ \nghost{{q}_{2}  } & \lstick{{q}_{2}  } &
\ghost{U_3}   & \qw & \\ \nghost{{q}_{3}  } & \lstick{{q}_{3}  } & \ghost{U_3} &
\qw & \\ \nghost{{q}_{4}  } & \lstick{{q}_{4}  } & \ghost{U_3} & \qw &}} =
\vcenter{ \Qcircuit @C=0.4em @R=0.2em @!R { \nghost{{q}_{0} } & \lstick{q_{1}  }
& \ctrl{3} & \qw & \qw & \qw & \qw & \qw & \qw & \qw & \ctrl{3} & \qw & \qw &
\qw & \qw & \qw & \qw & \qw \qw & \qw &\\ \nghost{{q}_{1} } &\lstick{q_{2}  }  &
\qw & \qw & \qw & \qw & \ctrl{2} & \qw & \qw & \qw & \qw & \qw & \qw & \qw &
\ctrl{2} & \qw & \qw & \qw & \qw &\\ \nghost{{q}_{2} } &\lstick{q_{3}  }  & \qw
& \qw & \ctrl{1} & \qw & \qw & \qw & \ctrl{1} & \qw & \qw & \qw & \ctrl{1} & \qw
& \qw & \qw & \ctrl{1} & \qw & \qw &\\ \nghost{{q}_{3} } &\lstick{q_{4}  } &
\targ & \gate{V^\dagger} & \targ & \gate{V} & \targ & \gate{V^\dagger} & \targ &
\gate{V} & \targ & \gate{V^\dagger} & \targ & \gate{V} & \targ &
\gate{V^\dagger} & \targ & \gate{V} & \qw &}} \] \caption{Quantum circuits for each of the unitaries $U_n$ as shown in figure \ref{Aeqn.gray_circuit}. For achieving an $N$-qubit C$^N$-P($\theta$) phase gate, the operators $V$ and $V^{\dagger}$ are the single qubit phases gates $P(\theta')$ and $P(-\theta ')$ respectively, where $\theta ' = \theta / 2^{N-1}$. } \label{Aeqn.gray_unitaries} \end{figure*}
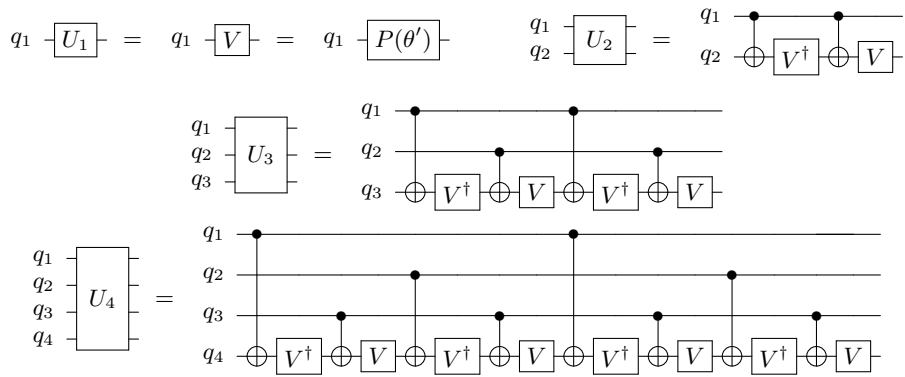

We note that the approach used in this study requires $O(2^N)$ CNOT gates, and thus is only
practical for small $N$ instances.  However, our approach did give shallower circuit depth and fewer gate counts than the native C$^N$-P($\theta$) implementation by Qiskit at the time of running (pre-Qiskit 2.0). For example,
at $N=5$ our transpilation for diffusion was 48 CNOTs as compared to Qiskit's {\tt MCPhase} gate which transpiled to 70.

\end{document}